\documentclass[fleqn,usenatbib]{mnras}
\usepackage{newtxtext,newtxmath}
\usepackage{orcidlink}
\DeclareUnicodeCharacter{2212}{-}
\DeclareUnicodeCharacter{2248}{=}
\usepackage[outercaption]{sidecap} 
\usepackage{wrapfig}
\usepackage{ulem}
\usepackage{xcolor}
\usepackage[switch, modulo]{lineno}
\usepackage[T1]{fontenc}
\hypersetup{
    colorlinks=true,
    linkcolor=blue,
    filecolor=magenta,      
    urlcolor=cyan,
    pdftitle={Overleaf Example},
    pdfpagemode=FullScreen,
    }
\usepackage{tablefootnote}

\DeclareRobustCommand{\VAN}[3]{#2}
\let\VANthebibliography\thebibliography
\def\thebibliography{\DeclareRobustCommand{\VAN}[3]{##3}\VANthebibliography}

\usepackage{graphicx}	% Including figure files
\usepackage{amsmath}	% Advanced maths commands

\title[Numerical quantification of the wind properties of cool main sequence stars]{Numerical quantification of the wind properties of cool main sequence stars}

\author[J. J. Chebly et al.]{
Judy J. Chebly \orcidlink{0000-0003-0695-6487}$^{1,2}$\thanks{E-mail: jchebly@aip.de},
Julián D. Alvarado-Gómez \orcidlink{0000-0001-5052-3473}$^{1}$,
Katja Poppenh\"{a}ger\orcidlink{0000-0003-1231-2194}$^{1,2}$,
Cecilia Garraffo\orcidlink{0000-0002-8791-6286}$^{3}$,
\\
$^{1}$Leibniz Institute for Astrophysics, An der Sternwarte 16, 14482, Potsdam, Germany\\
$^{2}$Institute of Physics and Astronomy, University of Potsdam, Potsdam-Golm, 14476, Germany\\
$^{3}$Harvard-Smithsonian Center for Astrophysics, 60 Garden Street, Cambridge, MA 02138, USA
}

\date{Accepted XXX. Received YYY; in original form ZZZ}

\pubyear{2023}

\begin{document}
\label{firstpage}
\pagerange{\pageref{firstpage}--\pageref{lastpage}}
\maketitle

\begin{abstract}
As a cool star evolves, it loses mass and angular momentum due to magnetized stellar winds which affect its rotational evolution. This change has consequences that range from the alteration of its activity to influences over the atmosphere of any orbiting planet. 
Despite their importance, observations constraining the properties of stellar winds in cool stars are extremely limited. Therefore, numerical simulations provide a valuable way to understand the structure and properties of these winds. In this work, we simulate the magnetized winds of 21 cool main-sequence stars (F-type to M-dwarfs), using a state-of-the-art 3D MHD code driven by observed large-scale magnetic field distributions. We perform a qualitative and quantitative characterization of our solutions, analyzing the dependencies between the driving conditions (e.g., spectral type, rotation, magnetic field strength) and the resulting stellar wind parameters (e.g., Alfv\'en surface size, mass loss rate, angular momentum loss rate, stellar wind speeds). We compare our models with the current observational knowledge on stellar winds in cool stars and explore the behaviour of the mass loss rate as a function of the Rossby number. Furthermore, our 3D models encompass the entire classical Habitable Zones (HZ) of all the stars in our sample. This allows us to provide the stellar wind dynamic pressure at both edges of the HZ and analyze the variations of this parameter across spectral type and orbital inclination. The results here presented could serve to inform future studies of stellar wind-magnetosphere interactions and stellar wind erosion of planetary atmospheres via ion escape processes.   

% \textbf{ max 250 words}
\end{abstract}


%%%%%%%%%%%%%%%%%%%%%%%%%%%%%%%%%%%%%%%%%%%%%%%%%%

%%%%%%%%%%%%%%%%% BODY OF PAPER %%%%%%%%%%%%%%%%%%

\section{Introduction}\label{sec:intro}
For many decades, scientists have known that the Sun has a mass outflow, which is most visible in the behavior of comet tails (e.g.,~\citealt{1957Obs....77..109B}). It has also been established that solar wind is a natural byproduct of the heating processes that produce the hot solar corona ($T~\sim~10^{6}$~K). As a result, all cool main-sequence stars ($M_{\bigstar}~\leqslant$~1.3~M$_\odot$) with analogous hot coronae, evidenced from their measured X-ray properties (\citealt{1995NIMPA.367..215S, 2000A&A...361..614P, Wright2011}), should have similar winds (\citealt{1958ApJ...128..664P}).
Magnetic fields are thought to play a key role as an energy source for the corona and the expanding solar atmosphere (e.g.,~\citealt{aschwanden2006physics, 2015RSPTA.37340256K, 2015RSPTA.37340262V}). Recent theories have shown that in addition to magnetic fields, wave dissipation (via turbulence) and magnetic reconnection could also play a role in energizing and shaping the spatial properties of the solar wind (see,~\citealt{2010LRSP....7....4O, 2012SSRv..172..145C, 2012SSRv..172...89H, 2015RSPTA.37340148C}).

Winds, even if relatively weak, play an important role in stellar evolution for stars of different spectral types causing the star to lose angular momentum and slow its rotation over time (\citealt{1967ApJ...148..217W, 1972ApJ...171..565S, 2012ApJ...754L..26M, 2013A&A...556A..36G,2015A&A...577A..98G, 2015A&A...577A..27Ja,2015A&A...577A..28Jb,2020A&A...635A.170A}). As a result, the magnetic activities that constitute the space weather (i.e., stellar winds, flares, coronal mass ejections)
 will decrease with age in low-mass stars
(\citealt{1972ApJ...171..565S, 1997ApJ...483..947G, 2005ApJ...622..680R, 2014MNRAS.438.1162V}).
These changes in the host star will also affect the evolution of planetary atmospheres and habitability (\citealt{2008DPS....40.1406T, 2014A&A...562A.116K, 2017ApJ...836L...3A}).

Direct measurements of the solar wind by spacecraft such as the Advanced Composition Explorer (ACE, \citealt{1998SSRv...86....1S, mccomas1998}), Ulysses \citep{2003GeoRL..30.1517M}, and Parker solar probe \citep{2021PhRvL.127y5101K} have improved our knowledge and understanding of its properties. 
On the other hand, detecting a solar-like wind emitted by another star has proven extremely challenging. This is not surprising, given how difficult it is to observe the solar wind remotely. The latter carries a very low mass loss rate ($\dot{M}_{\odot}  = \rm{ 2 \times 10^{-14}~\dot{M_{\odot}}~ {\rm yr^{-1}}}$, see \citealt{1977JGR....82..667F, Wood2004}), which implies relatively low densities (near the heliopause: $\sim$~0.002~cm$^{-3}$, \citealt{Gurnett2019}). Similarly, its high temperature and elevated ionization state, make it difficult to detect with simple imaging or spectroscopic techniques. 
As a result, properties such as the associated mass loss rates, angular momentum loss rates, and terminal velocities, crucial to understand stellar winds in low-mass stars, remain poorly constrained. 

Attempts to directly detect thermal radio emission from the plasma stream in cool stars have not yet led to any discovery (\citealt{lim1996, 1993ApJ...406..247D, 1997A&A...319..578V, 2000GeoRL..27..501G, 2014ApJ...788..112V, 2017A&A...599A.127F}).
Current radio telescopes are not optimized for this method; they can only detect winds much stronger than those from the Sun. Moreover, the coronae of these active stars are also radio sources, making it difficult to determine the exact source of the emission.
Nevertheless, this method has been able to establish upper limits for solar analogs of 1.3~$\times$~$10^{\rm -10}$~$\dot {M}_{\rm \odot}$~yr$^{\rm -1}$ (\citealt{2000GeoRL..27..501G, 2017A&A...599A.127F}).
Another proposed method for direct detection is to look for X-ray emission from nearby stars. As the star's winds propagate, they collide with the Local Interstellar Medium (ISM), forming "astrospheres" similar to the Sun's heliosphere \citep{Wood2004}. The charge exchange between the highly ionized stellar wind and the ISM produces X-ray photons with energies ranging from 453 to 701~eV. However, this method was unable to detect circumstellar charge exchange X-ray emission even from the nearest star, Proxima~Centauri \citep{2002ApJ...578..503W}. 

Similar to the charge exchange X-ray emission method, the Ly-$\rm \alpha$ absorption technique assumes the presence of the charge exchange phenomenon. In this case, however, we are interested in the neutral hydrogen wall formed at the astrospherical outer boundary by the interaction between the stellar wind and the ISM. This exchange has been detected as excess HI Ly-$\rm \alpha$ absorption in Hubble Space Telescope UV stellar spectra \citep{2014ASTRP...1...43L}. With nearly 30 measurements to date, spectroscopic analyses of the stellar HI lines have proven to be the best method to unambiguously detect and measure weak solar-like winds as well as some evolved cool stars \citep{Wood2021}. 

Using this method,\cite{2005ApJ...628L.143W} found evidence for some increase in $\dot{M}$ with magnetic activity, corresponding to a power-law relation in the form $\dot{M} \propto F_{\rm X}^{\rm 1.34 \pm 0.18}$ with $F_{\rm X} < 10^{\rm 6}$ $\rm erg~cm^{\rm -2}~s^{\rm -1}$. However, this relation does not seem to hold anymore for more active stars ($F_{\rm X} > 10^{\rm 6}$~$\rm erg$~cm$^{-2}$~s$^{-1}$), mainly M-dwarfs \citep{2005ApJ...628L.143W, 2014ApJ...781L..33W}. Recently, \cite{Wood2021} established a power law ($\dot{M} \propto F_{\rm X}^{\rm 0.77 \pm 0.04}$) between the $\dot{M}$ per unit surface area and the X-ray surface flux for coronal winds for a broader selection of stars, including G, K, and new $\dot{M}$ estimates for M-dwarfs. They found that the relation breaks even for stars with $F_{\rm X} < 10^{\rm 6}$ $\rm erg$~cm$^{-2}$~s$^{-1}$ (e.g.,~GJ~436, which has $F_{\rm X}$~=~4.9~$\times 10^{\rm 4}$~erg~cm$^{-2}$~s$^{\rm -1}$, where the $\dot{M}$ was estimated by using the planet as a probe for the stellar wind \citealt{Vidotto2017}) with the magnetic topology being a possible factor for the scatter.

While extremely useful, the search for astrospherical absorption is influenced by a number of critical factors. For instance, this method is strongly dependent on the relative velocity of the stellar rest frame and the ISM flow velocity ($V_{\rm ISM}$). As well as on the angle, $\theta$, between the upwind direction of the ISM flow and the line-of-sight to the star \citep{Wood2021}.
It also requires prior knowledge of the properties of the ISM such as the density and its ionization state (\citealt{Wood_2005}; \citealt{Redfield_2008}). Finally, its applicability is limited to relatively nearby stars ($\lesssim 15$~pc) due to the absorption of the ISM.

Due to the scarcity of observational data and associated limitations, numerical simulations can be used to improve our understanding of stellar winds. Models based on Alfv\'{e}n waves are more commonly used to simulate the stellar wind from stars other than the Sun \citep{2006ApJ...640L..75S}. This is because these waves are considered to be key mechanism for heating and accelerating the solar wind (\citealt{vanderHolst2014}; \citealt{2020SSRv..216..140V}).

In this study, we present a detailed numerical characterization of the stellar wind properties of cool main-sequence stars (early F to M-dwarfs) covering a range of rotation rates and magnetic field strengths. We compute steady-state stellar wind solutions using a state-of-the-art 3D MHD model and provide consistent qualitative and quantitative comparisons. Our goal is to better understand the different stellar wind properties as a function of the driving parameters, allowing us to explore the expected stellar wind conditions in the circumstellar region around planet-hosting stars.

This paper is organized as follows: Section \ref{Section 2} describes the numerical model and properties of the selected stellar sample. In Sect.~\ref{Section 3}, we present our numerical results, discuss the derived trends in the stellar wind properties, and compare our results with observations. This information  is then used to quantify the stellar wind conditions and explore their implications in the context of the classical habitable zone (HZ) around cool main-sequence stars. Conclusions and summary are provided in Sect.~\ref{Section 4}. 

\begin{figure}
 \centering
         \includegraphics[width=0.9\columnwidth]{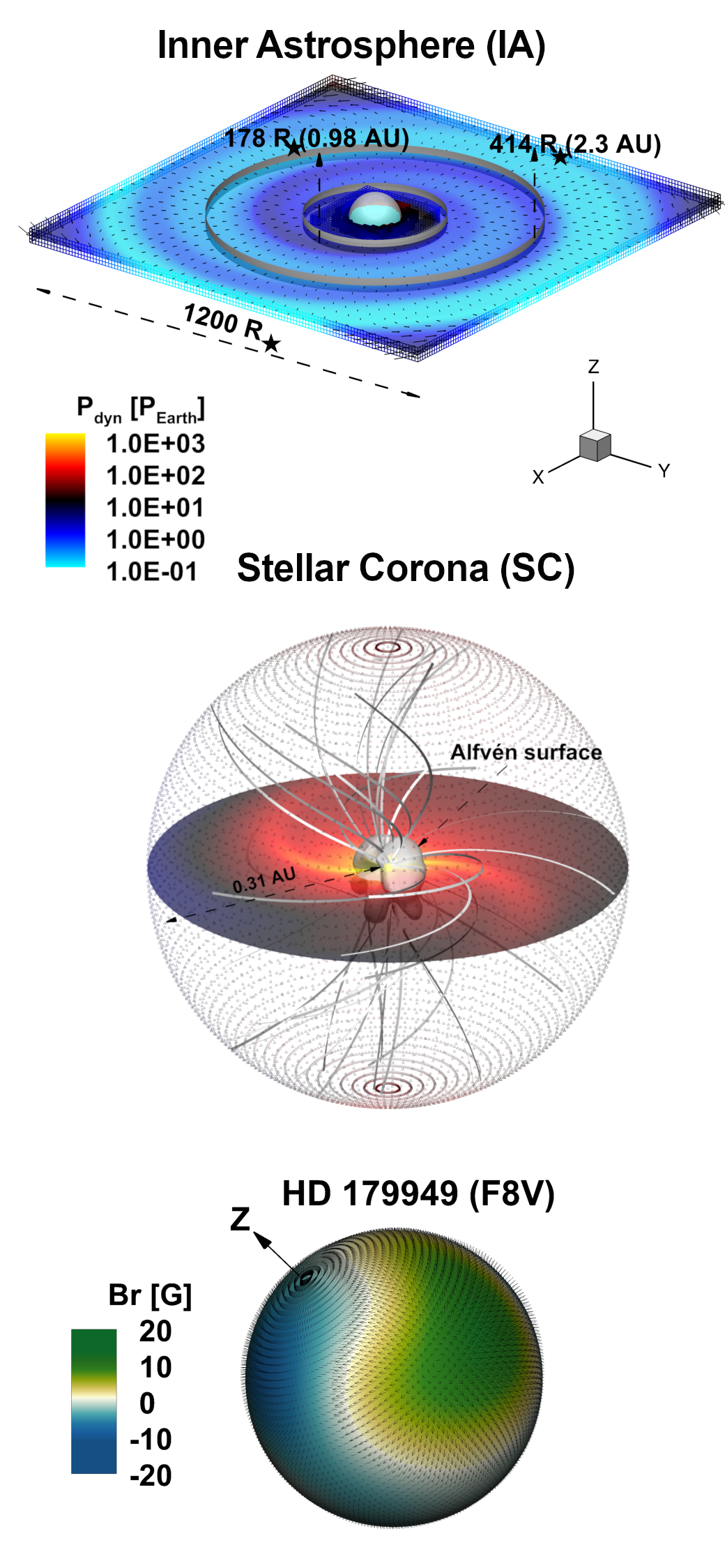}

         \caption{Simulated stellar wind environment for HD~179949 including multi-domain models. The bottom panel contains the surface field distribution (in Gauss) provided by the ZDI map and used to drive the AWSoM solution within the stellar corona (SC) domain (middle panel). The blue-green color bar represents the radial magnetic field strength on the stellar surface. Within SC, the gray iso-surface corresponds to the Alfv\'{e}n surface of the stellar wind (see Sect.~\ref{sec:3.1}). Selected magnetic field lines are shown in white. The steady-state solution is propagated from the coupling region (62~-~67R$_{\bigstar}$) to the entire Inner Astrosphere (IA) domain (1200~R$_{\bigstar}$ in each cartesian direction; upper panel). The central gray sphere in the top panel denotes the boundary of IA with the SC domain at 67~$R_{\bigstar}$. This domain contains the inner and outer edges of the habitable zone (gray circles). Color-coded (top and middle panel) is the wind dynamic pressure ($P_{\rm dyn}$~=~$\rho U^{\rm 2}$), normalized to the nominal Sun-Earth value ($\simeq$~1.5~nPa), visualized on the equatorial plane of both domains. The z-axis indicates the assumed stellar rotation axis of the star.}

         \label{figure1}
\end{figure}

\begin{figure*}
 \centering
         \includegraphics[width=0.9\textwidth]{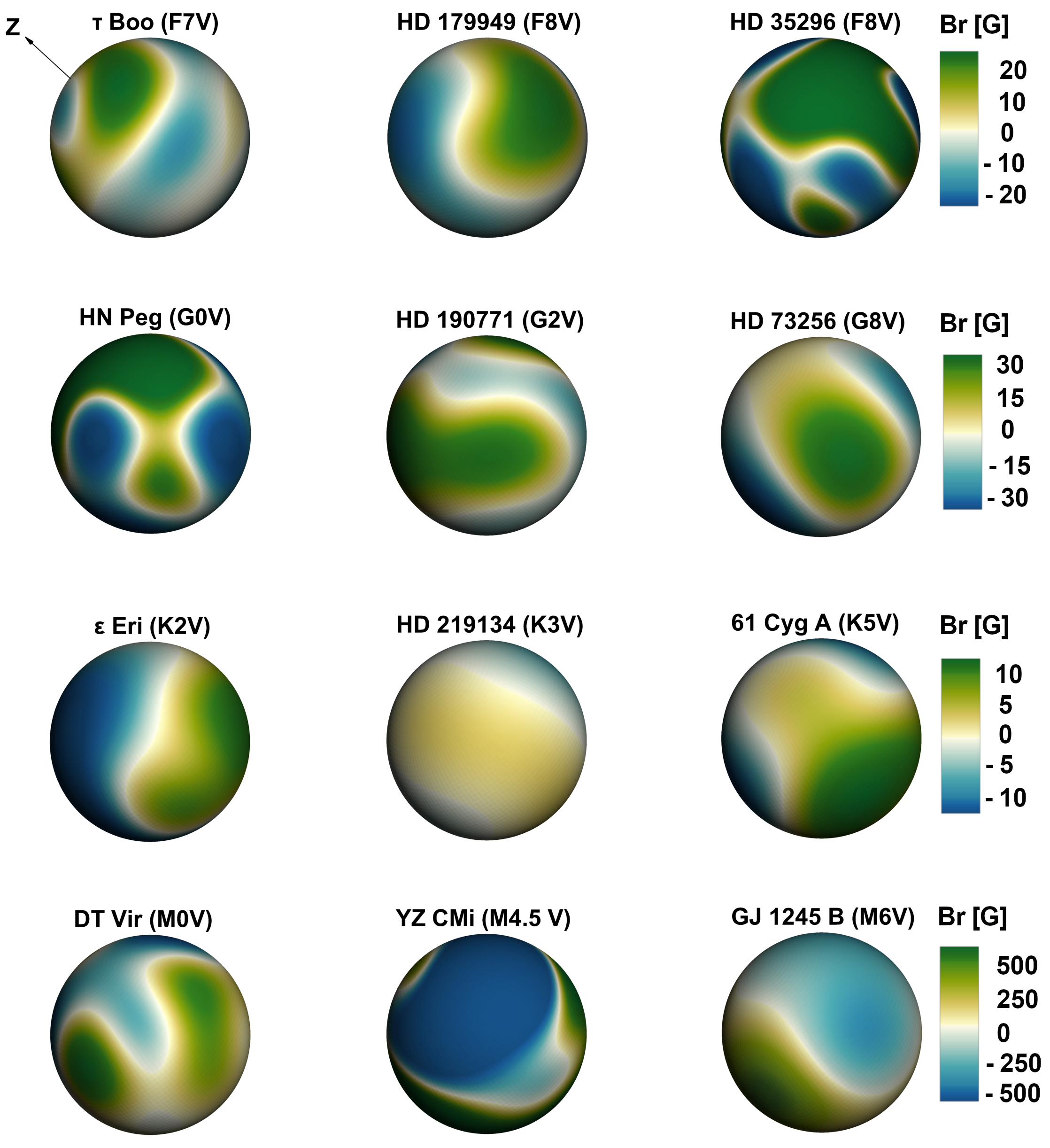}
         \caption{Examples of surface field distribution (in Gauss) of our sample stars retrieved from ZDI maps. The rows represent different spectral types going from late F-(top) to M-dwarfs (bottom) as indicated. The color-code represents the normalized radial magnetic field for a given row. The z-axis indicates the assumed stellar rotation axis for all the stars in the panel. The slowest rotation in our sample of 21 stars is HD~219134 (K3V,~$ P_{\rm rot}$~=~42.2~d), and the fastest is GJ~1245~B (M6V,~$P_{\rm rot}$~=~0.71~d). The radial magnetic field strength ranges from 5~G to 1.5~kG. 
}
\label{figure 2}
\end{figure*}

\begin{table*}
\centering
\caption{Fundamental parameters of our sample. Columns 1–8 list the star number, name, spectral type (SpT), stellar mass ($M_{\bigstar}$), stellar radius ($R_{\bigstar}$),  rotation period ($P_{\rm \scriptscriptstyle rot}$), effective temperature ($T_{\rm \scriptscriptstyle eff}$), and luminosity, respectively (\citealt{See2019} and references therein). }
\label{tab:1}
\begingroup
\setlength{\tabcolsep}{10pt} % Default value: 6pt
\renewcommand{\arraystretch}{1.5} % Default value: 1

\begin{tabular}{|p{16mm}|p{23mm}|p{10mm}||p{10mm}||p{10mm}||p{7mm}||p{13mm}||p{15mm}|}
 \hline % inserts single horizontal[3ex] \midrule \\
ID number &Star & SpT & $M_{\bigstar}$ [$M\odot$] & $R_{\bigstar}$ [$R\odot$] & $P_{\rm \scriptscriptstyle rot}$ [d] & \vtop{\hbox{\strut $T_{\rm eff}$} \hbox{\strut [K]}}  & \vtop{\hbox{\strut $L_{\rm \bigstar}$} \hbox{\strut [$L_{\rm \odot}$]}}  \\ 
  \hline % inserts single horizontal[3ex] \midrule

$1$ & $\tau$~Boo & F7V   & 1.34 & 1.46 & 3    & 6387 & 3.0  \\

$2$ & HD~179949 & F8V & 1.21 & 1.19 & 7.6  & 6168 & 1.80  \\

$3$ & HD 35296  & F8V & 1.06 & 1.1  & 3.48 & 6202 & 1.60  \\

$4$ & HN~Peg & G0V  & 1.1  & 1.04 & 4.55 & 5974 & 1.20  \\

$5$ & HD~190771 & G2V   & 1.06 & 1.01 & 8.8  & 5834 $\pm$ 50 & 0.99  \\

$6$ & TYC~1987-509-1& G7V   & 0.9  & 0.83 & 9.43 & 5550$^\dagger$  & 0.52 $\pm$ 0.03 \\

$7$ & HD~73256  & G8V  & 1.05 & 0.89 & 14   & 5480$^\dagger$  & 0.72 \\

$8$ & HD~130322 & K0V  & 0.79 & 0.83 & 26.1 & 5400 $^\dagger$ & 0.5  \\

$9$ & HD~6569 & K1V & 0.85 & 0.76 & 7.13 & 5170 & 0.36 $\pm$ 0.01   \\

$10$ & $\epsilon$~Eri & K2V  & 0.85 & 0.72 & 11 & 5125 $\pm$ 87 & 0.3 $\pm$ 0.06 \\

$11$ & HD~189733 & K2V  & 0.82 & 0.76 & 12.5 & 4939 & 0.34  \\

$12$ & HD~219134& K3V  & 0.81 & 0.78 & 42.2 & 4835 $^\dagger$ & 0.27 \\

$13$ & TYC 6878-0195-1& K4V & 0.65$^\ddag$ & 0.64$^\ddag$ & 5.72  & 4600$^\dagger$ & 0.8 $\pm$ 0.32 \\

$14$ & 61~Cyg~A & K5V & 0.66 & 0.62 & 34.2 & 4655 $^\dagger$ & 0.15  \\

$15$ & HIP 12545 & K6V & 0.58$^\ddag$  & 0.57$^\ddag$  & 4.83 & 4300 $^\dagger$ & 0.4 $\pm$ 0.06  \\

$16$ & TYC 6349-0200-1 & K7V& 0.54$^\ddag$ & 0.54$^\ddag$  & 3.39 & 4100$^\dagger$ & 0.3 $\pm$ 0.02 \\

$17$ & DT Vir & M0V  & 0.59 & 0.53 & 2.85 & 3850$^\dagger$ & 0.055  \\

$18$ & GJ~205 & M1.5V & 0.63 & 0.55 & 33.6 & 3690$^\dagger$& 0.061$\pm$ 0.006 \\

$19$ & EV~Lac & M3.5V & 0.32 & 0.3  & 4.37 & 3267 & 0.013  \\

$20$ & YZ~CMi & M4.5V & 0.32 & 0.29 & 2.77 & 3129 & 0.012 \\

$21$ & GJ~1245~B & M6V & 0.12 & 0.14 & 0.71 & 3030$^\S$ & 0.0016  \\
\hline

\end{tabular}%}
\endgroup
\begin{flushleft}
\footnotesize{$^\ddag$ Reference from \cite{10.1093/mnras/stu728}}\\
\footnotesize{$^\dagger$ Reference from \cite{2013ApJS..208....9P}}\\
\footnotesize{$^\S$ Reference from \cite{2013A&A...557A..67V}}
\end{flushleft}
\end{table*}

\section{Model description} \label{Section 2}

We simulate stellar winds in cool main-sequence stars using the state-of-the-art Space Weather Modeling Framework (SWMF;~\citealt{2013ApJ...764...23S, vanderHolst2014, 2018LRSP...15....4G}). The SWMF is a set of physics-based models (from the solar corona to the outer edge of the heliosphere) that can be run independently or in conjunction with each other \citep{Toth2012}. 
This model uses the numerical schemes of the Block Adaptive Tree Solar Roe-Type Upwind Scheme (BATS-R-US;~\citealt{1999JCoPh.154..284P}) MHD solver.
For a detailed description of the model, see \citet{2021JSWSC..11...42G}.
The multi-domain solution starts with a calculation using the Solar/Stellar Corona (SC) module which incorporates the Alfv\'{e}n Wave Solar Model (AWSoM; \citealt{vanderHolst2014}). This module provides a description of the coronal structure and the stellar wind acceleration region. The simulation is then coupled to a second module known as the Inner Heliosphere/Astrosphere\footnote{This module is formally labeled IH within the SWMF, but since we are working with low-mass main sequence stars, we will refer to it as the Inner Astrosphere (IA) domain.} (IA). In this way, it is possible to propagate the stellar wind solution up to Earth's orbit and beyond. The model has been extensively validated and updated employing remote sensing as well as in-situ solar data (e.g., \citealt{2017ApJ...845...98O,2019ApJ...887...83S,2019ApJ...872L..18V}).

AWSoM is driven by photospheric magnetic field data, which is normally available for the Sun in the form of synoptic magnetograms \citep{2014SoPh..289..769R}.
A potential field source surface method is used to calculate the initial magnetic field (more details in the following section). This information is used by AWSoM to account for heating and radiative cooling effects, as well as the Poynting flux entering the corona, and empirical turbulent dissipation length scales. With the interplay between the magnetic field distribution, the extrapolation of the potential field, and the thermodynamic properties, the model solves the non-ideal magnetohydrodynamic (MHD) equations for the mass conservation, magnetic field induction, energy (coronal heating), and momentum (acceleration of the stellar wind). These last two aspects are controlled by Alfv\'{e}n waves propagating along and against the magnetic field lines (depending on the polarity of the field). In the momentum equation, the heat and acceleration contributions are coupled by an additional term for the total pressure and a source term in the energy equation. The numerical implementation is described in detail in \citet{vanderHolst2014}.
Once these conditions are provided, the simulation evolves all equations locally until a global steady-state solution is reached.

\subsection{Simulation parameters and setup}

In our work, we  apply the SWMF/AWSoM model to main-sequence F, G, K, and M-type stars by assuming that their stellar winds are driven by the same process as the solar wind. 
We analyze the properties of the stellar wind by a coupled simulation covering the region of the stellar corona (SC,~spherical) and the resulting structure within the inner astrosphere (IA,~cartesian). Figure~\ref{figure1} illustrates the coupling procedure in one of our models. This coupling was necessary only in the case of F, G, and K stars, in order to completely cover the habitable zones (HZ)\footnote{The range of orbits around a star in which an Earth-like planet can sustain liquid water on its surface.}, which are larger and farther away from the star. Parameters such as stellar radius ($R_{\bigstar}$), mass ($M_{\bigstar}$), and rotation period ($P_{\rm rot}$), are also taken into account in the simulations. We followed the approach in \cite{Kopparapu2014} in order to determine the optimistic HZs boundaries of each star in our sample.

\subsubsection{Simulation domain}

The star is positioned in the center of the SC spherical domain. The radial coordinate in SC ranges from 1.05~$R_{\bigstar}$ to 67~$R_{\bigstar}$, except for M-dwarfs, where it extends to 250~$R_{\bigstar}$. The choice of the outer edge value of the SC domain was chosen in a way to obtain both edges of the HZ in one domain. The habitable zones limits were calculated using \cite{Kopparapu2014} approach and the reported measured $L_{\bigstar}$ and $T_{\rm eff}$ for each star in our sample (see Table~\ref{tab:1}). As will be discussed in Sect.~\ref{Section 3}, in the case of M-dwarfs, the extension had to be performed in order to cover the entire Alfv\'{e}n surface~(AS)\footnote{This structure sets the boundary between the escaping wind and the magnetically coupled outflows that do not carry angular momentum away from the star.}, while keeping the default parameters for AWSoM fixed (see Sect. \ref{sec:iParams}). The domain uses a radially stretched grid with the cartesian z-axis aligned with the rotation axis. 
The cell sizes in the meridional ($\phi$) and azimuthal ($\theta$) directions are fixed at $\sim$~$2.8 ^\circ$. The total number of cells in the SC domain is $\sim$ $8 \times 10^{5}$. \\

The steady-state solutions obtained within the SC module are then used as inner boundary conditions for the IA component. An overlap of 5~$R_{\bigstar}$ (from 62~$R_{\bigstar}$ to 67~$R_{\bigstar}$) is used in the coupling procedure between the two domains for F, G, and K stars (more details on the necessity of the overlap when coupling between domains can be found in \citealt{Toth2005}). The IA is a cube that extends from $62 ~R_{\bigstar}$ to $600~R_{\bigstar}$ in each cartesian component. Adaptive Mesh refinement~(AMR) is performed within IA, with the smallest grid cell size of $\sim$~1.17~$R_{\bigstar}$ increasing up to 9.37~$R_{\bigstar}$ with a total of 3.9~million cells. As the simulation evolves, the stellar wind solution is advected from SC into the larger IA domain where the local conditions are calculated in the ideal MHD regime.

\subsubsection{Magnetic boundary conditions}

In the initial condition of the simulation, observations are used to set the radial component of the magnetic field $B_{\rm r}$~[G] anchored at the base of the wind (at the inner boundary). As mentioned earlier, a finite potential field extrapolation procedure is carried out to obtain the initial configuration of the magnetic field throughout SC \citep{2011ApJ...732..102T}. This procedure requires setting an outer boundary (source surface, $r_{\rm s}$), beyond which the magnetic field can be considered to be purely radial and force-free. The magnetic field can therefore be described as a gradient of a scalar potential and determined by solving Laplace's equation in the domain. For the simulations discussed here, we set $r_{\rm s}$ at $45\%$ of the SC domain size for F, G, and K stars, and $70 \%$ for M-dwarfs. While the choice of this parameter does not alter significantly the converged solutions, it can modify the required run time of each model to achieve convergence. Therefore, our selection was done to guarantee convergence to the steady-state in a comparable number of iterations between all spectral types. 

The stellar magnetic field as reconstructed from Zeeman Doppler Imaging (ZDI)\footnote{A tomographic imaging technique that allows the reconstruction of the large-scale magnetic field (strength and polarity at the star’s surface from a series of polarized spectra (see e.g., \citealt{Donati2006, 2008MNRAS.390..567M, 2009MNRAS.398.1383F,
2015A&A...582A..38A,2016A&A...585A..77H, Kochukhov202}).}, is used as the inner boundary condition of SC (Fig.~\ref{figure 2}). Therefore, the resulting wind solutions are more realistic than models based on simplified/idealized field geometries \citep{Judy2021}. Although the reconstructed maps provide the distribution of vector magnetic fields, we use only the radial component of the observed surface field.
The magnetogram is then converted into a series of spherical harmonic coefficients with a resolution similar to that of the original map.
The order of the spherical harmonics should be chosen so that artifacts such as the "ringing" effect do not appear in the solution \citep{2011ApJ...732..102T}. In our models, we performed the spherical harmonics expansion up to $l_{\rm max}$~=~5. 

\subsubsection{Input parameters}\label{sec:iParams}

After we set the initial conditions, we define several parameters for the inner boundary. 
In order to reduce the  degree of freedom of the parameter set, we only modify the parameters related to the properties of the stars, such as mass, rotation period, and radius.
As for the other parameters, we implement the same values that are commonly used in the solar case (\citealt{vanderHolst2014,2019ApJ...887...83S}). 
The Poynting flux ($ S/B_{\rm \bigstar} = 1.1\times 10^{6}$~J~m$^{-2}$~s$^{-1}$~T) is a parameter that determines the amount of wave energy provided at the base of a coronal magnetic field line.
The other parameter is the proportionality constant that controls the dissipation of Alfv\'{e}n wave energy into the coronal plasma and is also known as the correlation length of Alfv\'{e}n waves ($L\rm{_{\rm \perp} = 1.5\times 10^{5}}~m~\sqrt{T}$). We use the values given in \cite{2013ApJ...764...23S} to define the base temperature ($T_{\rm o} = 2 \times 10^{6} K$) and the base density ($n\rm{_{\rm o} = 2\times 10^{11}}  cm^{-3}$). 

We note that the choice of these parameters will affect the simulation results, as reported in several studies that followed different approaches (e.g.,~\citealt{Boro2020, 2023SpWea..2103262J}). Recently, \cite{2023SpWea..2103262J} performed a global sensitivity analysis to quantify the contributions of model parameter uncertainty to the variance of solar wind speed and density at 1~au. They found that the most important parameters were the phostospheric magnetic field strength, $S/B_{\rm \bigstar}$, and $L\rm{_{\rm \perp}}$. Furthermore, in \cite{Boro2020}, an increase in the mass loss rate~($\dot{M}_{\bigstar}$), and angular momentum loss rate~($\dot{J}_{\bigstar}$) was reported when $S/B_{\rm \bigstar}$ is increased from the solar value to $2.0\times 10^{6}$~J~m$^{-2}$~s$^{-1}$~T), which is expected because $S/B_{\rm \bigstar}$ drives the energy of the Alfv\'en wave, resulting in higher $\dot{M}_{\bigstar}$ and $\dot{J}_{\bigstar}$.

In this work, however, we are interested in isolating the expected dependencies with the relevant stellar properties (e.g., mass, radius, rotation period, photospheric magnetic field) which can only be analyzed consistently if the AWSoM related parameters are kept fixed between spectral types. Moreover, as will be discussed in detail in Sect.~\ref{sec:Mdot-Ro}, the results obtained using the standard AWSoM settings are either consistent with current stellar wind observational constraints for different types of stars or the apparent differences can be understood in terms of other physical factors or assumptions made in the observations. For these reasons, we have chosen not to alter these parameters in this study, which also reduces the degrees of freedom in our models.

\subsection{The sample of stars}
Our investigation is focused on main sequence stars, with effective temperatures ranging from 6500~K down to 3030~K, and masses $M_{\rm \bigstar} < 1.34~M_{\rm \odot}$ (spectral types F to M). All of these stars are either fully or partially convective. We use a sample of 21 stars whose large-scale photospheric magnetic fields were reconstructed with ZDI (\citealt{2019ApJ...876..118S} and references therein). Some of these stars were observed at different epochs. In this case, the ZDI map with the best phase coverage, signal-to-noise ratio, and most spectra used in the reconstruction was chosen. 
The sample includes radial magnetic field strengths in the ZDI reconstruction between $5$~G and $1.5$~kG corresponding to HD~130322~(K0V) and EV~Lac~(M3.5V), respectively. Spectral types range from F7~($\tau$~Boo,~$M_{\bigstar} = 1.34~M_{\odot}$, $R_{\bigstar} = 1.46~R_{\odot}$) to M6~(GJ~1245~B,~$M_{\bigstar} = 0.12~M_{\odot}$, $R_{\bigstar} = 0.14~R_{\odot}$).
The rotation periods vary between fractions of a day to tens of days, with GJ~1245~B~(M6V) having the shortest rotation period ($P_{\scriptscriptstyle \rm rot }$~=~0.71~d) and HD~219134~(K3V) the longest one ($P_{\scriptscriptstyle \rm rot }$~=~42.2~d). Table~\ref{tab:1} contains the complete list of the sample stars and a summary of the stellar properties incorporated in our models. 

\begin{figure}
 \centering
 \includegraphics[width=0.9\columnwidth]{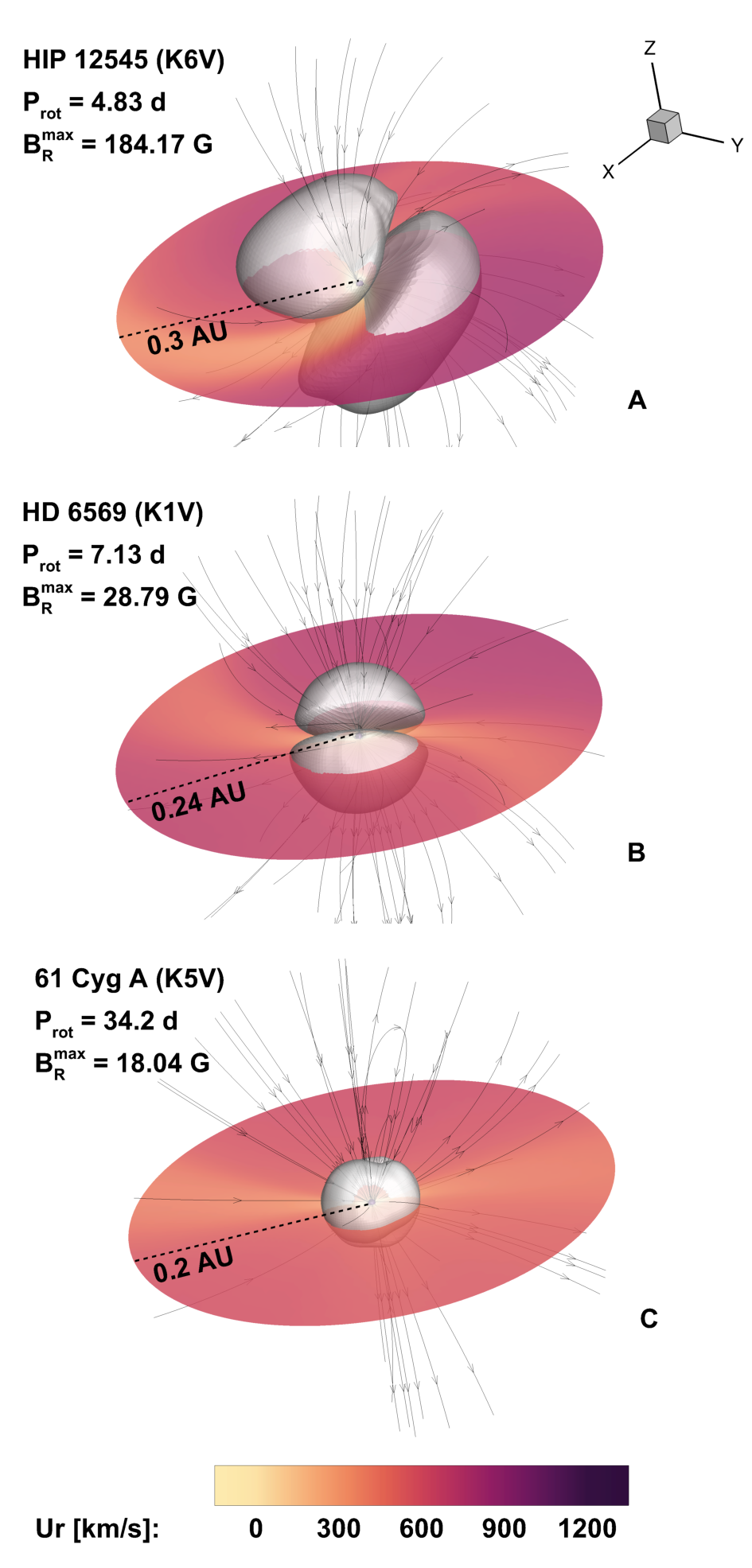}
         \caption{Simulation results in the SC domain for 3 K stars: HD~12545 (panel A), HD~6569 (panel~B), 61~Cyg~A (panel~C) driven by ZDI magnetic field maps. All panels contain the projection onto the equatorial plane ($z$~=~0) of the radial wind velocity ($U_{\rm r}$).The translucent gray shade denotes the Alfv\'{e}n surface calculated from the steady-state solution. The corresponding color scale $U_{\rm r}$ is preserved among the different panels. Selected 3D magnetic field lines are shown in black. The absolute size of the SC domain is indicated in each case.
}
\label{figure3}
\end{figure}

\begin{figure*}
 \centering
         \includegraphics[width=1\textwidth]{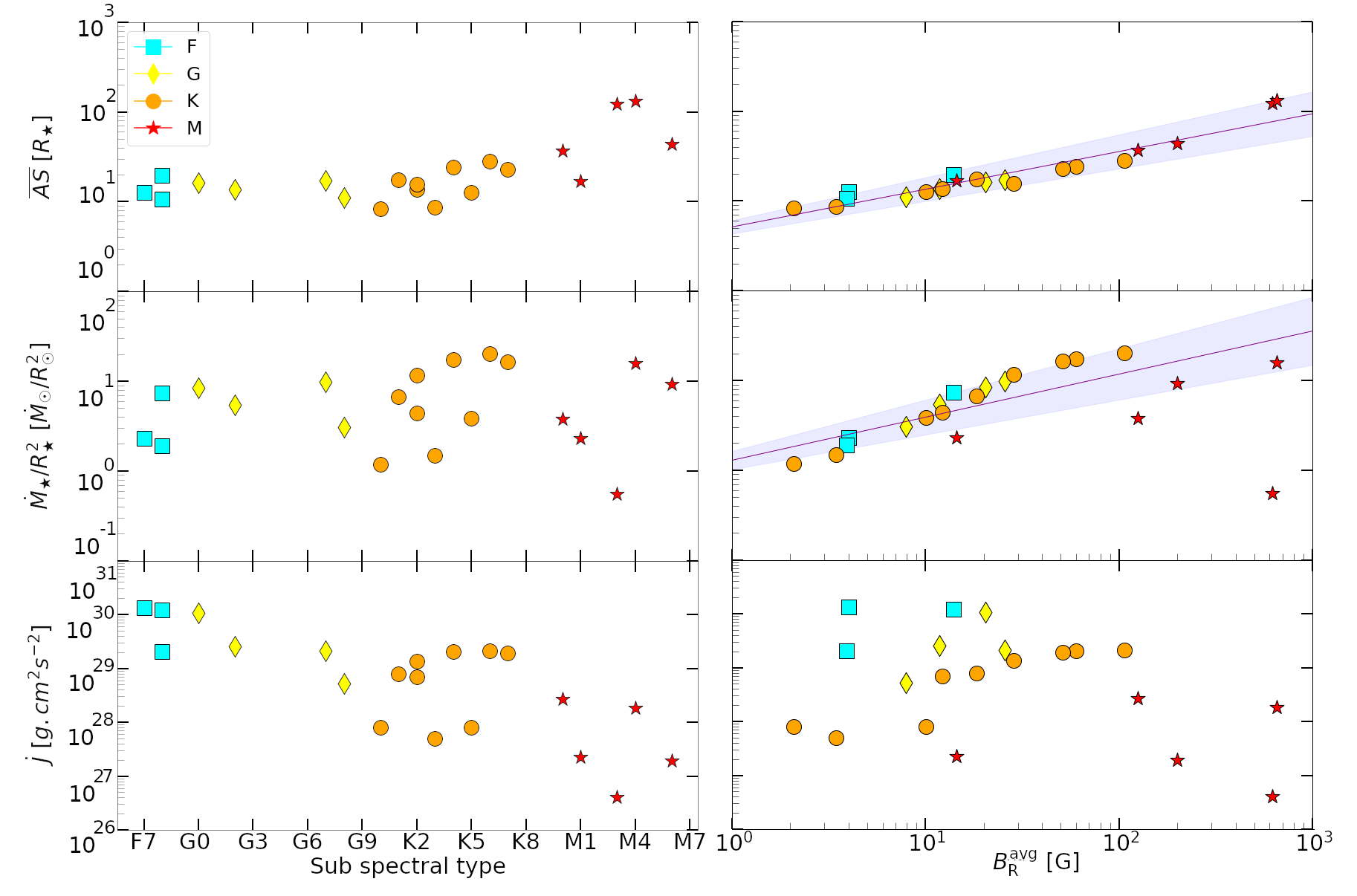}
         \caption{Simulated average Alfv\'en surface ($\overline {AS}$, top), mass loss rate per unit surface area ($\dot{M}_{\bigstar}/R^{\rm 2}_{\rm \bigstar}$, middle), and angular momentum loss rate ($\dot{J}$, bottom) as a function of the spectral type (left) and the surface-average radial magnetic field ($B_{\rm R}^{\rm avg}$, right). The mass loss rate is expressed in units of the average solar values ($\dot{M}_{\rm \odot} = 2 \times 10^{\rm -14}~M_{\rm \odot}~{\rm yr^{\rm -1}} =  1.265 \times 10^{\rm 12}~{\rm g~s^{-1}}$), normalized to the surface area of each star ($A_{\bigstar} = 4\pi R_{\bigstar}^{2}$). Individual points denote the results of each simulation presented in Sect. \ref{Section 3}, Table \ref{tab:2}. The different symbols and colors represent the spectral types (F, cyan/squares; G, yellow/diamonds; K, orange/circles; M, red/star). The purple line and shaded area represent the fitted power-law with its uncertainties. \label{fig4}}
\end{figure*}

\section{Results \& Discussion} \label{Section 3}
\subsection{The effect of star properties on the wind structure}\label{sec:3.1}

The Alfv\'{e}n surface (AS) is defined by the collection of points in the 3D space that fulfils the Alfv\'{e}n radius criterion\footnote{The Alfv\'{e}n radius ($R_{\rm A}$) is defined as the distance around a star at which the kinetic energy density of the stellar wind equals the energy density of the astrospheric magnetic field.}. Numerically, it is determined by finding the surface for which the wind velocity reaches the local Alfv\'{e}n velocity, $v_{A} = B/\sqrt{4 {\pi \rho}}$, where $B$ and $\rho$ are the local magnetic field and plasma density, respectively. The Alfv\'{e}n surface can be interpreted as the lever arm of the wind torque --the "position" at which the torque acts to change the angular rotation of the star\footnote{In other words, the angular momentum per unit mass within the stellar wind can be computed as if there were solid body rotation, at an angular velocity $\Omega_{\bigstar} = 2\pi/P_{\rm rot}$, out as far as the Alfv\'{e}n surface.}. The Alfv\'{e}n Surface is used in numerical models to characterize ($\dot{M}_{\bigstar}$) and ($\dot{J}_{\bigstar}$) (e.g.,~\citealt{2015MNRAS.449.4117V,Boro2020, 2015ApJ...813...40G, 2016A&A...594A..95A}).
We compute $\dot{J}_{\bigstar}$ by performing a scalar flow rate integration over the AS and another one over a closed spherical surface (S) beyond the AS to determine $\dot{M}_{\bigstar}$:

\vspace{- 0.1 cm}
\begin{equation}
\dot{M}_{\bigstar} = \int_{S} \rho (\textbf{u} \cdot \rm \textbf{dA})\label{eq1}
\end{equation}
\vspace{- 0.25 cm}
\begin{equation}
\dot{J}_{\bigstar} = \int_{AS} \rm \Omega \rho \rm R^{2} \sin^{2} \theta  (\textbf{u} \cdot \rm \textbf{dA})\label{eq2}
\end{equation}

\noindent Here $\dot{J}_{\bigstar}$ is the component of the change in angular momentum in the direction of the axis of rotation. The distance to the Alfv\'{e}n surface is represented by $R$. The angle between the lever arm and the rotation axis is denoted by $\theta$, which depends on the shape/orientation of the AS with respect to the rotation axis (and accounted for in the surface integral). The stellar angular velocity is represented by $\Omega = 2 \rm \pi/ P_{\scriptscriptstyle \rm rot}$. The surface element is denoted by \textbf{dA}.

Figure \ref{figure3} shows the AS of the stellar wind, with plasma streamers along with the equatorial section flooded with the wind velocity ($U_{\rm r}$) for three K stars in our sample (HIP~12545, HD~6569, 61~Cyg~A). If we compare two stars with similar $P_{\rm \scriptscriptstyle rot}$ but different $B_{\rm R}^{\rm max}$, we can clearly see that the size of AS increases with increasing magnetic field strength. This is a direct consequence of the dependence of the Alfv\'{e}n velocity on these quantities (Eq. \ref{eq1}) and the distance from the star at which the Alfv\'{e}n velocity is exceeded by the wind. For instance, for very active stars with stronger magnetic fields, the expected coronal Alfv\'{e}n velocity is greater than for less active stars, increasing the radial distance that the wind velocity must travel to reach the Alfv\'{e}n velocity. The associated Alfv\'{e}n surface has a characteristic two-lobe configuration (Fig.~\ref{figure3}, gray translucent area), with average sizes of 27~$R_{\bigstar}$, 18~$R_{\rm \bigstar}$ and 13~$\rm R_{\bigstar}$ for HIP~12545, HD~6569, and 61~Cyg~A, respectively (see Table \ref{tab:2}). 

When we compare two stars with similar magnetic field strengths but different $ P_{\rm \scriptscriptstyle rot}$ (see Fig.~\ref{figure3}, panels B and C), the change in AS size is not as dramatic. The rotation period has primarily a geometric effect on the resulting AS. The Alfv\'{e}n surface assumes a different tilt angle in all three cases. This tilt is mainly connected to the open magnetic field flux distribution on the star's surface \citep{Garraffo2015}. We also notice in Fig.~\ref{figure3} that the stellar wind distribution is mainly bipolar with a relatively fast component reaching up to $\sim$\,$891$~km~s$^{−1}$ for HIP~12545~$\sim$, $702$~km~s$^{−1}$ for HD~6569, and $\sim$\,$593$~km~s$^{−1}$ for 61~Cyg~A. In section \ref{sec:obs-sim} we will discuss further the relation between the wind velocity with regard to $P_{\rm rot}$ and $B_{\rm R}$. 

Figure~\ref{fig4} shows the $\dot{M}_{\bigstar}$, $\dot{J}_{\bigstar}$, $\overline {AS}$ as estimated by the previously described method, against the sub-spectral type of our star sample (left column) and the average radial magnetic field strength ($B_{\rm R}^{\rm avg}$,~right column).  Similar relations have been obtained for the maximum radial magnetic field strength and are presented in Appendix~\ref{sec:appendix}. The average Alfv\'en surface size was calculated by performing a mean integral over the radius at each point of the 3D AS. The extracted quantities are represented by different colors and symbols for each spectral type (F, G, K, and M).

As expected, the AS increases as we move toward more magnetically active stars (Fig. \ref{fig4}, top-right panel). From our simulations, we were able to establish a relation between AS and $B_{\rm R}^{\rm avg}$ using the bootstrap technique (1000 realizations) to find the mean of the slope and the intercept along with their uncertainties. We use this approach to determine all relations from our simulations.
The relation is as follows: 

\begin{equation}
   \log \overline{AS_{\rm R}} = (0.42 \pm 0.06) \log B_{\rm R}^{\rm avg} + (0.71 \pm 0.07)\label{eq3}
\end{equation}

\noindent Our simulated steady-state $\dot{M}_{\bigstar}$ show a scatter within the range [0.5~$\dot{M}_{\odot}$/$R_{\scriptscriptstyle \odot}^{2}$, 30~$\dot{M}_{\odot}$/$R_{\scriptscriptstyle \odot}^{2}$], which is comparable to that estimated from the observed Ly$\rm \alpha$ absorption method of G, K, and M-dwarfs in \cite{Wood2021}.
The variations in $\dot{M}_{\bigstar}$ are related to differences in the strength and topology of the magnetic field driving the simulations (see \citealt{2016A&A...594A..95A, 2016MNRAS.455L..52V, 2022MNRAS.510.5226E}), as well as to the Alfv\'{e}n wave energy transfer to the corona and wind implemented in the model (\citealt{2020A&A...635A.178B, 2023SpWea..2103262J}). For this reason, we tried to isolate the effects introduced by the star (e.g.,~$M_{\bigstar}$, $R_{\bigstar}$, ${P_{\rm rot}}$, magnetic field strength) over the ones from the Alfv\'{e}n wave heating (i.e.,~$n_{\rm o}$, $T_{\rm o}$, $S/B_{\bigstar}$, $L_{ \perp}$). 

In terms of mass loss rate, stronger winds are expected to be generated by stronger magnetic fields (see~Fig.~\ref{fig4}) implying that the winds are either faster or denser. This interplay determines $\dot{M}_{\bigstar}$ (Eq.~\ref{eq1}), which increases with increasing magnetic field strength regardless of spectral type. We see a common increase for F, G, K, and M-dwarfs (excluding EV~Lac) in the saturated and unsaturated regime that can be defined from the simulations as follows:

\begin{equation}
    \log \dot{M}_{\rm \bigstar} / R^{\rm 2}_{\bigstar} = (0.48 \pm 0.09) \log B_{\rm R}^{\rm avg} + (0.11 \pm 0.10)\label{equ4}
\end{equation}

\noindent On the other hand, we observe a slightly different behavior for M-dwarfs, whose $\dot{M}_{\bigstar}$ and $\dot{J}_{\bigstar}$ values tend to be lower. As discussed by \citet{2018ApJ...862...90G}, the magnetic field complexity could also affect $\dot{M}_{\bigstar}$ for a given field strength. We consider this possibility in the following section. Note that, as has been shown in previous stellar wind studies of M-dwarfs (e.g.,~\citealt{2017ApJ...843L..33G, 2021MNRAS.504.1511K, 2022ApJ...928..147A}), modifications to the base AWSoM parameters (either in terms of the Poynting flux or the Alfv\'en wave correlation length) would lead to strong variations in $\dot{M}_{\bigstar}$. This would permit placing the M-dwarfs along the general trend of the other spectral types in particular, the $\dot{M}_{\bigstar}$ value obtained for the star with the strongest $B_{\rm R}$ in our sample (EV~Lac). While these modifications have physical motivations behind them (i.e. increased chromospheric activity, stronger surface magnetic fields), in most regards, they remain unconstrained observationally. Furthermore, the values we obtain in our fiducial AWSoM models are still within the range of observational estimates available for this spectral type (see~Sect.~\ref{sec:obs-sim}), with the added benefit of minimizing the degrees of freedom and isolating the effects of the stellar parameters on the results. 

Similarly, we see a large scatter of $\dot{J}_{\bigstar}$ with respect to the spectral type (Fig. \ref{fig4}, bottom left column), ranging from $\rm{10^{26}~{\rm g~cm^{2}~s^{-2}}}$ to $\rm{10^{31}~{\rm g~cm^{2}~s^{-2}}}$. This range is within the expected $\dot{J}_{\bigstar}$ values estimated for cool stars with the lowest value corresponding to M-dwarfs (\citealt{See2019_magneticfield} and references therein). The maximum $\dot{J}_{\bigstar}$ values reached in our simulations are comparable to $\dot{J}_{\odot}$ reached at solar minimum and maximum ($7\times 10^{\rm 30}$ and $ 10 \times 10^{\rm 30}$~$\rm g.cm^{2}s^{-2}$, \citealt{2018ApJ...864..125F}; \citealt{Boro2020}). 
We note that this is the only parameter for which we have retained units in absolute values (as is commonly done in solar/stellar wind studies; see \citealt{2010ApJ...723L..64C}; \citealt{Garraffo2015}; \citealt{2018ApJ...864..125F}). Using absolute units, we expect a decrease in $\dot{J}_{\bigstar}$ as we move from F to M-dwarfs, since $\dot{J}_{\bigstar}$ is a function of $R^{\rm2}$ (Eq.~\ref{eq2}). The scatter around this trend is dominated by the relatively small $\dot{M}_{\bigstar}$ values, the distribution of $\Omega_{\bigstar}$ in our sample (variations up to a factor of 5),  and the equatorial AS size where the maximum torque is applied ($\sin{\theta}$ in Eq. \ref{eq2}). 

We also note that the sample is biased toward weaker magnetic field strengths. To better estimate how the magnetic field affects the properties of the stellar winds, we need a larger sample, not only in terms of stellar properties but also with stellar wind constraints such as $\dot{M}_{\bigstar}$. The latter is so far the only stellar wind observable parameter for which comparisons can be made. For this reason, we will focus on the behavior of the $\dot{M}_{\bigstar}$ as a function of different stellar properties in the following sections of the analysis.

% \section{Discussion}
\subsection{Stellar mass-loss rate and complexity}

\begin{figure*}
 \centering
         \includegraphics[width=1\textwidth]{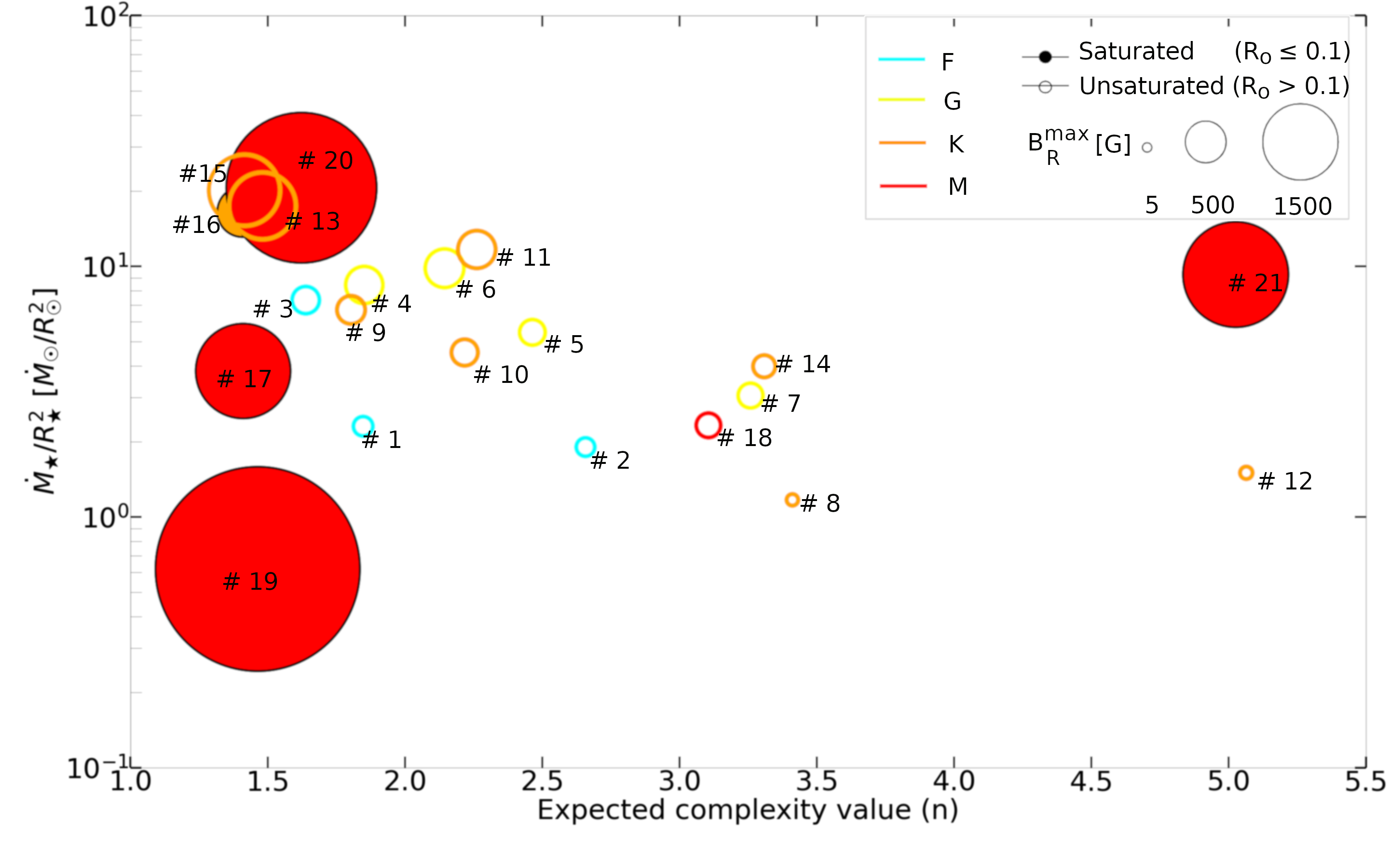}
         \caption{Simulated mass loss rate ($\dot{M}_{\rm \bigstar}/ R^{\rm 2}_{\rm \bigstar}$) versus the expected complexity value ($n$, \citealt{2018ApJ...862...90G}) given the Rossby number ($R_{\rm o}$) of each star. The mass loss rate is expressed in solar units, normalized by the unit surface area of each star. Individual points denote the results of each simulation presented in Sect. 3 (Table \ref{tab:2}). The different colors (cyan, orange, yellow, red) represent the different spectral types respectively (F, G, K, and M). The filled circles and the empty ones represent stars in the saturated and unsaturated regimes, respectively. The symbol size denotes the maximum magnetic field strength ($B_{\scriptscriptstyle \rm R}^{\rm max}$) in each case. The numbers refer to the ID of the star in our sample. The complexity value n = 1 corresponds to the dipole, and n increases as we move to the right.}\label{fig5}
\end{figure*}

Coronal X-ray luminosity is a good indicator of the level of magnetic activity of a star and the amount of material heated to $10^6$~K temperatures. The dependence of magnetic activity on dynamo action (i.e.,~dynamo number $D = R_{\rm o}^{-2}$, \citealt{2020LRSP...17....4C}) has led a number of authors to use the Rossby number to characterize stellar activity, for a wide range of stellar types \citep{Wright2011}.
The Rossby number is defined as $R_{\rm o} = P_{\rm rot}$/$ \tau_{\rm c}$, where $P_{\rm rot}$ is the stellar rotation period and $ \tau_{\rm c}$  is the convective turnover time (\citealt{1984ApJ...279..763N, Jordan1991, 2011ApJ...743...48W}). We adopted the approach of \cite{2018MNRAS.479.2351W} to calculate $\tau_{\rm c}$. In this case, the latter is only a function of the stellar mass~($M_{\rm \bigstar}$):
\hspace{-1cm}
\begin{equation}
\log \tau_{\rm c} = 2.33 - 1.50 (M_{\bigstar}/M_{\odot}) + 0.31 (M_{\bigstar}/M_{\odot})^{2} \label{eq5}
\end{equation}

\noindent As it was mentioned in Sect.~\ref{sec:intro}, the study of \cite{Wood2021} suggests that coronal activity increases with $\dot{M}_{\bigstar}$. The overall increase in $\dot M_{\bigstar}$ with X-ray flux $F_{\rm X}$ ($\dot M_{\bigstar} \propto F_{\rm X}^{0.77\pm0.04}$), is most likely due to their dependence on magnetic field strength (see Sect.~\ref{sec:3.1}). However, they report a scatter of about two orders of magnitude of $\dot M_{\bigstar}$ around the trend line. This suggests that coronal activity and spectral type alone do not determine wind properties. The geometry of the magnetic field may also play a role.

The correlation between $\dot M_{\bigstar}$ and magnetic complexity has already been suggested by \cite{2015ApJ...813...40G}, which could in principle contribute to the scatter in (\citealt{Wood2021}, Fig.~10).
The large-scale distribution of the magnetic field on the stellar surface is mainly determined by the rotation period and the mass of the star, namely $R_{\rm o}$ 
(\citealt{2018ApJ...862...90G, 2010MNRAS.407.2269M, 2013A&A...549L...5G}). 
The Rossby number was used to determine the complexity function in \citet{2018ApJ...862...90G}, which was able to reproduce the bimodal rotational morphology observed in young open clusters (OCs). The complexity function of \citet{2018ApJ...862...90G} is defined as 
\begin{equation}
n = \frac{a}{R_{\rm o}}+ 1 +  bR_{\rm o}\mbox{.}\label{eq6} 
\end{equation}

\noindent The constant 1 reflects a pure dipole. The coefficients a~=~0.02 and b~=~2 are determined from observations of OCs. The first term is derived from the ZDI map observation of stars with different spectral types and rotation periods. The third term is motivated by Kepler's observations of old stars (\citealt{2016Natur.529..181V,2019ApJ...872..128V, 2016Natur.529..181V, 2019ApJ...872..128V}).

We emphasize that the complexity number ($n$), estimated from Eq.~\ref{eq6}, differs from the complexity derived from the ZDI maps themselves (e.g., \citealt{2022ApJ...941L...8G}). The complexity number from $R_{\rm o}$ is expected to be higher. This is due to the fact that many of the small-scale details of the magnetic field are not captured by ZDI. 

We expect to lose even more information about the complexity of the field given that the ZDI maps are not really available to the community (apart from the published images). Image-to-data transformation techniques (which we applied to extract the relevant magnetic field information from the published maps) can lead to some losses of information, both spatially and in magnetic field resolution. These vary depending on the grid and the projection used to present the ZDI reconstructions (i.e., Mercator, flattened-polar, Mollweide). Using the star's raw ZDI map would prevent these issues and would aid with the reproducibility of the simulation results.

Finally, note that the expected complexity is also independent of the spherical harmonic expansion order used to parse the ZDI information to the simulations. The obtained $R_{\rm o}$ and $n$ values for each star in our sample are listed in Table \ref{tab:2}.

Figure \ref{fig5} shows the behaviour of coronal activity and $\dot{M}_{\bigstar}$  with respect to the expected magnetic field complexity (\textit{n}). The coronal activity is denoted by full and empty symbols corresponding to saturated and unsaturated stars, respectively. We consider stars with $R_{\rm o} \le 0.1$ in the saturated regime and stars with $R_{\rm o} > 0.1$ in the unsaturated regime based on X-ray observations (\citealt{2003A&A...397..147P, 2011ApJ...743...48W, Wright2018, Stelzer2016, 2016Natur.535..526W}). The colors correspond to the different spectral types, whereas the numbers indicate the ID of each star in our sample. The symbol size represents the maximum radial magnetic field strength of each star extracted from the ZDI observations.

We anticipate seeing a trend in which the $\dot{M}_{\bigstar}$ decreases as the magnetic field complexity increases (leading to an increment of closed loops on the stellar corona), for stars in saturated and unsaturated regimes. For instance, $\epsilon$~Eri~($\#10 $, $B_{\scriptscriptstyle \rm R}^{\scriptscriptstyle \rm max}= 25$ G,~n~=~2.21724) has an $\dot{M}_{\rm \bigstar}$~=~4.53~$\dot{M}_{\rm \odot}$/$R_{\odot}^{\rm 2}$ lower than HD~6569~($\#9$,~$B_{\scriptscriptstyle  \rm R}^{\scriptscriptstyle  \rm max}=~\rm 29$ G,~n~=~1.80346~) with $\dot{M}_{\bigstar}$~=~6.70~$\dot{M}_{\rm \odot}$/$R_{\odot}^{\rm 2}$. This is also true for $\tau$~Boo and HD~179949 where $\tau$~Boo ($\#1 $, $B_{\scriptscriptstyle \rm R}^{\scriptscriptstyle \rm max}= 14$~G,~n~=~1.84728, $\dot{M}_{\bigstar}$~=~2.30~$\dot{M}_{\rm \odot}$/$R_{\odot}^{\rm 2}$) has a higher $\dot{M}_{\bigstar}$ compared to HD~179949~($\#2 $, $B_{\scriptscriptstyle \rm R}^{\scriptscriptstyle \rm max}= 12$~G,~n~=~2.65746, $\dot{M}_{\bigstar}$~=~1.90~$\dot{M}_{\rm \odot}$/$R_{\odot}^{\rm 2}$).

We also noticed that as we go to more active stars, like in the case of M-dwarfs, the field strength starts to dominate over the complexity in terms of contribution to the $\dot{M}_{\bigstar}$. For example, GJ~1245~B~($\#21$,~$B_{\scriptscriptstyle  \rm R}^{\scriptscriptstyle  \rm max}=~404$~G~,~n~=~5.02602~, $\dot{M}_{\bigstar}$~=~9.27~$\dot M_{\rm \odot}$/$R_{\odot}^{\rm 2}$) has an $\dot{M}_{\bigstar}$ higher than DT~Vir even though the complexity of the former is almost 5 times higher (DT~Vir,~$\#17$,~$B_{\scriptscriptstyle \rm R}^{\scriptscriptstyle \rm max}=~327$~G,~n~=~1.41024, $\dot{M}_{\bigstar}$~=~3.81~$\dot{M}_{\rm \odot}$/$R_{\odot}^{\rm 2}$). However, in order to better understand the contribution of the complexity in $\dot{M}_{\bigstar}$, we will need to run simulations for a wider range of stars with sufficiently high resolution of the driving magnetic field to capture directly the complexity of the field (and not estimate it from a scaling relation as it was performed~here).

Moreover, our results show that whenever we have a case in which the star properties ($M_{\bigstar}$, $R_{\bigstar}$, and $P_{\rm rot}$), magnetic field strength and complexity are comparable, we end up with similar $\dot{M}_{\bigstar}$. This will be the case of TYC~6878-0195-1~($\#13$,~$B_{\scriptscriptstyle  \rm R}^{\scriptscriptstyle  \rm max}=~162$~G,~n~=~1.48069, $\dot{M}_{\bigstar}$=~17.42~$\dot M_{\rm \odot}$/$R_{\odot}^{\rm 2}$) and HIP~12545~($\#~15$,~$B_{\scriptscriptstyle  \rm R}^{\scriptscriptstyle  \rm max}=~184$~G,~n~=~1.41505, $\dot{M}_{\bigstar}$~=~20.11~$\dot M_{\rm \odot}$/$R_{\odot}^{\rm 2}$). 

Furthermore, two stars with similar coronal activity with respect to X-ray flux, i.e.,~EV~Lac and YZ~CMi~($F_{\scriptscriptstyle \rm X} \approx 10^{7}$~$\rm ergs$~$ \rm cm^{-2}\rm s^{-1}$), but with slightly different magnetic field complexity, result in different wind properties: respectively $\dot{M}_{\rm \bigstar} = \rm 0.62$ $\dot M_{\rm \odot}$/$R_{\odot}^{\rm 2}$, and $\dot{M}_{\rm \bigstar} = \rm 20.57$ $\dot{M}_{\rm \odot}$/$R_{\odot}^{\rm 2}$.
A similar situation occurs when two stars have a comparable field complexity but different coronal activity i.e.,~YZ~CMi and GJ~205~($\#18$, $\dot{M}_{\rm \bigstar} = \rm 2.32$ $\dot{M}_{\rm \odot}$/$R_{\odot}^{\rm 2}$, $F_{\scriptscriptstyle \rm X} \approx 10^{5}$~$\rm ergs$~cm$^{-2}\rm s^{-1}$). 

The lowest $\dot{M}_{\bigstar}$ corresponds to the saturated M-dwarf EV~Lac~($\#$19), which has the strongest $B_{\scriptscriptstyle \rm R}$ (1517~G) and one of the simplest complexities in our sample (n~=~1.46331). The low complexity of the field means that the wind is dominated by open field lines, leading to very high wind velocities in the standard AWSoM model, but with a very low density, which in turn leads to small $\dot{M}_{\bigstar}$ values. We remind the reader that the base density of the stellar wind is fixed at the stellar surface and is the same for all the stars in the sample (Sect.~\ref{sec:iParams}).

\begin{figure*}
 \centering
         \includegraphics[width=1.\textwidth]{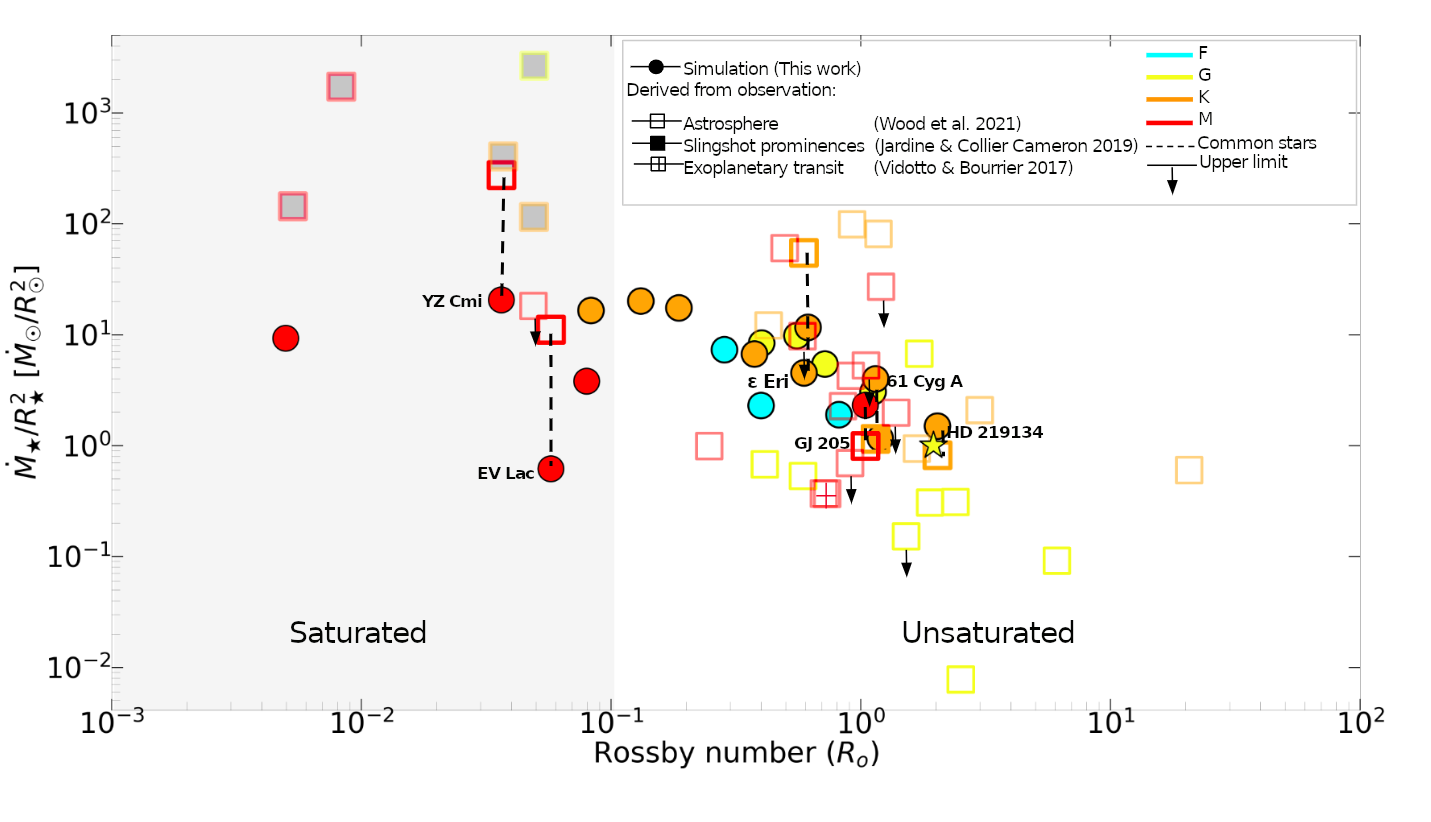}
         \caption{Numerical results of the stellar mass loss rate ($\dot{M}_{\bigstar}$/$R^{\rm 2}_{\bigstar}$, circles), astrosphere stellar mass loss rate (squares,~\citealt{Wood2021}), slingshot prominences mass loss rate (diamonds,~\citealt{2019MNRAS.482.2853J}), and absorption during an exoplanetary transit (plus within a square, \citealt{Vidotto2017}) against the Rossby number ($R_{\rm o}$). Colors illustrate the different spectral types: cyan (F), yellow (G), orange (K), and red (M). The Sun is represented by a star symbol. Dashed lines connect the common stars between our sample and the ones with estimated $\dot{M}_{\bigstar}/R^{\rm 2}_{\bigstar}$ values by the astrospheric \textcolor{red}{Ly-}$\alpha$ absorption technique. Black arrows pointing downward correspond to the upper limits given by observations.}
        
\label{fig:fig6}
\end{figure*}

\begin{figure*}
     \includegraphics[width=0.9\textwidth]{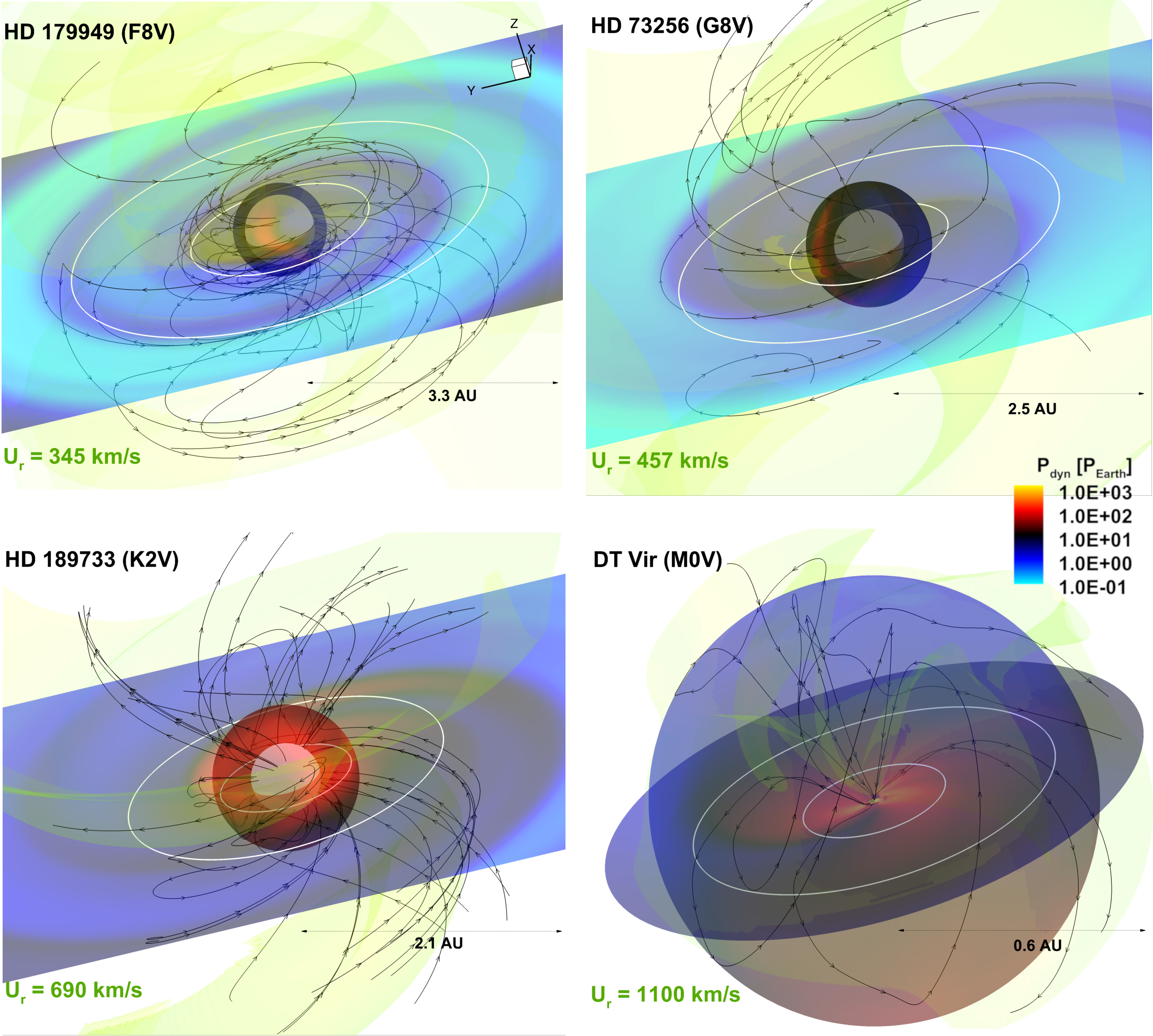}
    \caption{Simulated stellar wind environment for HD~179949, HD~73256, HD~189733, and DT~Vir. Multi-domain models for HD~179949, HD~73256, HD~189733 are shown. The steady-state solution of the multi-domain cases is propagated from the coupling region (62~$−$~67~$R_{\bigstar}$) to the entire IA domain (1200~$\rm R_{\bigstar}$ in each cartesian direction). The steady-state solution of DT~Vir is shown in the spherical domain (SC). The green iso-surface represents the averaged wind velocity at 1~au for F, G, and K, as for M-dwarfs it represents the highest averaged velocity. Color-coded is the wind dynamic pressure ($P_{\rm dyn} = \rm {\rho U^{2}}$ ) normalized to the nominal Sun-Earth value ($\sim$~1.5~nPa), visualized on the equatorial plane of both domains and on a translucent sphere (at R~=~0.5~au). Selected magnetic field lines are shown in black. The 2 white circles represent the optimistic habitable zone boundaries. The white translucent sphere represents the coupling region (67~$R_{\bigstar}$) between the SC and IH domains in the case of F, G, and K stars. }
    \label{fig:fig7}
\end{figure*}

\subsection{Stellar wind mass-loss rate and Rossby number}\label{sec:Mdot-Ro}

\begin{table*}
\Large
\centering
\caption{A summary of the resulting stellar wind properties with their corresponding driving parameters. Columns 1–11, list the star number, name, mass-loss rate per unit surface area$^\dagger$ ($\dot{{M}}_{\bigstar}/\rm {R}^2_{\bigstar}$), angular momentum loss rate ($\dot{{J}}_{\bigstar}$), average terminal velocity ($\overline{U_{\rm \scriptscriptstyle R}^{\rm \scriptscriptstyle 1 au}}$), average terminal velocity ($\overline{U_{\rm \scriptscriptstyle R}^{\rm \scriptscriptstyle T}}$), average Alfv\'en surface radius ($\overline{AS}$), absolute maximum surface radial magnetic field ($|B_{\rm R}|^{\rm max}$), average surface radial magnetic field ($B_{\rm R}^{\rm avg}$), Rossby number$^\ddag$ ($R_{\rm o}$), and the expected complexity number$^\S$ ($n$).}\label{tab:2}
\begingroup

\setlength{\tabcolsep}{10pt} % Default value: 6pt
\renewcommand{\arraystretch}{1.5} % Default value: 1
\resizebox{0.9\textwidth}{!}{%
\begin{tabular}{p{20mm}|p{35mm}|p{10mm}||p{14mm}||p{14mm}|p{10mm}|p{7mm}|p{11mm}|p{11mm}|p{15mm}|p{15mm}}

  \hline % inserts single horizontal[3ex] \midrule \\
ID number & Name &
  $\dot{M}_{\bigstar}/ R^2_{\bigstar}$ & \vtop{\hbox{\strut $\dot{J}_{\bigstar}$} \hbox{\strut  [$g.cm^{2}.s^{-1}$]}} 
   & \vtop{\hbox{\strut  $\overline{U_{\rm \scriptscriptstyle R}^{\rm \scriptscriptstyle 1 au}}$ } \hbox{\strut  [km $s^{-1}$]}} 
  & \vtop{\hbox{\strut  $\overline{U_{\rm \scriptscriptstyle R}^{\rm \scriptscriptstyle T}}$ } \hbox{\strut  [km $s^{-1}$]}} 
 & $\overline{AS}$ [R$_{\bigstar}$] & \vtop{\hbox{\strut $|B_{\rm R}|^{\rm max}$} \hbox{\strut  [G]}}&  \vtop{\hbox{\strut $B_{\rm R}^{\rm avg}$} \hbox{\strut [G]}}  & $R_{\rm o}$ & \textit{n} \\
    \hline % inserts single horizontal[3ex] \midrule 
$1$ & $\rm \tau$ Boo  & 2.30 & 1.35E+30 &  320 & 332 & 13 & 14 & 4.01 & 0.39855& 1.84728 \\

$2$ & HD~179949       & 1.90 & 2.07E+29 &345
&  356  & 11 &12 & 3.91 & 0.81648 & 2.65746\\

$3$ & HD~35296        & 7.33 & 1.20E+30 &  270
 &289  & 20 & 27 & 13.94 & 0.28396 & 1.63835\\

$4$ & HN~Peg          & 8.42 & 1.07E+30 & 545
  & 549 & 16 & 50 & 20.48 & 0.40079 & 1.85148 \\

$5$ & HD~190771       & 5.45  & 2.58E+29 &  431
&432  & 14 & 24 & 11.75 & 0.71806 & 2.46397\\

$6$ & TYC~1987-509-1  & 9.82& 2.11E+29 & 547
&530  &  17 & 54 & 25.63 & 0.55388& 2.14387 \\

$7$ & HD~73256        & 3.04 & 5.18E+28 & 457
&458& 11 & 22& 7.91  & 1.12036& 3.25857 \\

$8$ & HD~130322       & 1.17 & 8.01E+27 & 436
 &440&  9 & 5& 2.08 & 1.19721&  3.41113\\

$9$ & HD~6569         & 6.70 & 7.88E+28 &  704
 &702   & 18 & 29& 18.41  &0.37507 & 1.80346 \\

$10$ & $\epsilon$~Eri  & 4.53 & 6.89E+28& 554
 &553  & 14  & 25 &  12.26 & 0.59172 & 2.21724\\

$11$ & HD~189733       &11.67  &1.34E+29&  535
 &534   & 16 & 51 & 28.43 & 0.61443& 2.26141\\

$12$ & HD~219134       & 1.50 &4.97E+27& 425
  &429    & 9 & 6 &  3.44  & 2.02732  &  5.06451\\

$13$ & TYC 6878-0195-1 &17.42 &2.08E+29 & 734
 &692 &24 & 162& 60.14 & 0.18682& 1.48069 \\

$14$ & 61~Cyg~A        & 3.98& 7.98E+27 & 609 &593 &  13 & 18 &  10.07  & 1.14548&  3.30842 \\

$15$ & HIP 12545       & 20.11 &2.13E+29 &906
 &891 & 28 & 184 & 106.62 & 0.13145 & 1.41505 \\

$16$ & TYC 6349-0200-1 & 16.55  & 1.92E+29& 
657 & 642&23 & 93&  51.36 & 0.08314 & 1.40684 \\

$17$ & DT Vir          & 3.81 & 2.67E+28& ----& 1102 & 37 & 327 &  125.08 & 0.07979 & 1.41024\\

$18$ & GJ~205          & 2.32 & 2.28E+27 &  ----&690 &  17  & 22 &  14.41 & 1.04304& 3.10525\\

$19$ & EV~Lac          & 0.62 & 4.10E+26 &---- &3675  & 122& 1517 & 620.21  & 0.05738 &  1.46331 \\

$20$ & YZ~CMi          & 20.57  & 1.81E+28& ----&1709   & 132 & 822 &  655.66  & 0.03637&  1.62264\\

$21$ & GJ~1245~B       &  9.27& 1.90E+27& ---- &1164   & 44 & 404 &  200.43  & 0.00498 & 5.02602\\
\hline
\end{tabular}%}
}
\endgroup
\begin{flushleft}
\footnotesize{$^\dagger$ Normalized to solar units ($\dot{M}_{\odot}/{R}^2_{\odot} = 1.0$).}\\
\footnotesize{$^\ddag$ Predicted by the empirical model of \cite{Wright2011}.}\\
\end{flushleft}
\end{table*}

Using the results of our stellar winds models, we can study how the $\dot{M}_{\bigstar}$ changes as a function of the Rossby number ($R_{\rm o}$). The Rossby number is a useful quantity because it not only removes the dependence on spectral type, but also relates the rotation period to magnetic field strength, complexity, and even stellar coronal activity.
The latter is also important because cool stars exhibit a well-defined behavior between $L_{\rm X}$ (or $F_{\rm X}$) and $R_{\rm o}$ (saturated and unsaturated regimes). Thus, if we analyze $\dot{M}_{\bigstar}$ using this parameter, we can see (to some extent) all dependencies simultaneously.

Figure \ref{fig:fig6} shows the stellar mass-loss rate per unit surface area ($\dot{M}_{\bigstar}/R^{\rm 2}_{\bigstar}$) as a function of the Rossby number ($R_{\rm o}$). The circles show our 3D MHD numerical results, while the empty, filled, and the plus sign within a square corresponds to observational estimates of astrospheres \citep{Wood2021}, slingshot prominences \citep{2019MNRAS.482.2853J}, and absorption during an exoplanetary transit \citep{Vidotto2017}, respectively. We use the same method as for the simulated stars (Eq.~\ref{eq5}) to calculate the $R_{\rm o}$ of stars with constraints on their mass loss rate.
Spectral types are indicated by different colors: cyan~(F), yellow~(G), orange~(K), and red~(M). The Sun is represented by a yellow star symbol. Dashed lines connect the common stars in our models and the observations. In this section, we will focus only on the resulting $\dot{M}_{\bigstar}$ from the numerical results.

As was mentioned earlier, our 3D MHD simulated $\dot{M}_{\bigstar}$ values are in the same range as the $\dot{M}_{\bigstar}$ estimates from the Ly-$\alpha$ astrospheric absorption method. Note that since we are only simulating steady-state stellar winds, our comparison is mostly focused on the steady mass loss $\dot{M}_{\bigstar}$ (filled squares and squares with a plus sign). As such, it is not surprising that our $\dot{M}_{\bigstar}$ values appear 1~-~2 orders of magnitude below the estimates associated with sporadic mass loss events such as slingshot prominences in very active stars in the saturated regime (filled squares,~\citealt{2019MNRAS.482.2853J}).

Based on the relation between $F_{\rm X}$ and $R_{\rm o}$ \citep{2018MNRAS.479.2351W}, and the broad correlation observed between $\dot{M}_{\bigstar}$ and $F_{\rm X}$ \citep{Wood2021}, we expect to see traces of a two-part trend (albeit with significant scatter) between $\dot{M}_{\bigstar}$ and $R_{\rm o}$: a flat or saturated part that is independent of stellar rotation ($R_{\rm o} \lesssim 0.1$, rapidly rotating stars), and a power law showing that the stellar wind mass loss rate decreases with increasing $R_{\rm o}$ ($R_{\rm o} > 0.1$, slowly rotating stars). 

For stars in the unsaturated regime, we do see a trend in which $\dot{M}_{\bigstar}$ increases with decreasing $R_{\rm o}$. The relationship between $\dot{M}_{\bigstar}$ and $R_{\rm o}$ retrieved from our simulations is 
\vspace{-0.05cm}
\begin{equation}
\log \dot{M}_{\bigstar}/R_{\bigstar}^2 = (-1.13\pm 0.23) \log R_{\rm o} + (0.50\pm 0.07) \label{eq7}.     
\end{equation}

\noindent The majority of the $\dot{M}_{\bigstar}$ derived from observation appears to follow the established relationship $\dot{M}_{\bigstar}$--$R_{\rm o}$, with some scatter within the error range. We do, however, notice four outliers, including three K stars and one G star. The K stars with the high $\dot{M}_{\bigstar}$ correspond to the binary 70~Oph~A~(K0V) and 70~Oph~B~(K5V). As for the $3^{\rm \scriptscriptstyle rd}$ K star and the G star, they correspond to evolved stars: $\delta$~Eri~(K0IV,~$\dot{M}_{\bigstar}~=~0.6~\dot{M}_{\rm \odot}$/$R_{\odot}^{\rm 2}$, $R {\rm o}\,\sim~21$) and DK~UMA~(G4III-I, $\dot{M}_{\bigstar}~=~0.0077~\dot{M}_{\rm \odot}$/$R_{\odot}^{\rm 2}$, $R_{\rm o}$~$\sim$~2.51). 

We do not expect evolved stars to follow the same trend as unsaturated main sequence stars because their winds might be generated from a different mechanism (such as pulsations, see \citealt{2021LRSP...18....3V}). 
As for 70~Oph~A and B, we do not have much insight into their eruptive activity levels in order to rule out whether or not the $\dot{M}$ inferred from the astropsheric technique was influenced by slingshot prominences or CME activity.

As can be seen in Fig.~\ref{fig:fig6}, our numerical results in this region are essentially bracketed by the observations for which the $R_{\rm o}$ reaches larger values. The largest Rossby number from our star sample corresponds to HD~219134 (K3V, $R_{\rm o} = 2.02732$), which is comparable to the accepted solar value. Since our models use ZDI maps as inner boundary conditions to simulate stellar winds, this implies that extending our numerical models to even larger $R_{\rm o}$ would be very challenging as those ZDI reconstructions would require prohibitively long observing campaigns. 

While we have limited data points, we see that for objects with $R_{\rm o} \lesssim $~0.15, we do not obtain larger numerical values $\dot{M}_{\bigstar}$ even when the magnetic field strengths increase dramatically. For example, in the case of YZ~CMi~($B^{\rm max}_{\rm R}$ =~822~G, $\dot{M}_{\bigstar}/R_{\bigstar}^{\rm 2}$~=~20.57~$\dot M_{\rm \odot}$/$R_{\odot}^{\rm 2}$) and GJ~1245~B~($B^{\rm max}_{\rm R}$~=~404~G,~$\dot{M}_{\bigstar}$~=~9.27~$\dot M_{\rm \odot}$/$R_{\odot}^{\rm 2}$).
All stars on the left-hand side of Fig. \ref{fig:fig6} lie beneath the maximum $\dot{M}_{\bigstar}$ value obtained for YZ~CMi~($B^{\rm max}_{\rm R}$ =~822~G,~$\dot{M}_{\bigstar}/R_{\bigstar}^{\rm 2}$ =~20.57~$\dot M_{\rm \odot}/R_{\odot}^{\rm 2}$).
This is true even when $R_{\rm o}$ varies by more than one dex, magnetic field strength by factors of 100, and the expected complexity number by $\sim$\,4. 

These results indicate that the contribution from the steady wind will only account for a small fraction of the $\dot{M}_{\bigstar}$ budget in the case of very active stars. Furthermore, the obtained behaviour hints of a possible saturation of the steady-state stellar wind contribution to $\dot{M}_{\bigstar}$, while the star could still lose significant mass through other mechanisms such as slingshot prominences or CME activity due to flares among others.  

According to \cite{Villarreal2018} and references therein, cool stars can support prominences if their magnetospheres are within the centrifugal regime (i.e. $R_{\rm K} < R_{\rm A}$, where $R_{\rm K} = \sqrt[3]{GM_{\bigstar}/\Omega_{\bigstar}^2}$ is the co-rotation radius). They provide estimates for the prominence masses ($m_{\rm p}$) and the ejection time-scales ($t_{\rm p}$) for a sample of cool stars. According to their analysis, DT~Vir would have $m_{\rm p} =  1.5 \times 10^{\rm 15}$~g and $t_{\rm p}$~=~0.1~d, while the values for GJ~1245~B would be $m_{\rm p} =  4.4 \times 10^{\rm 14}$~g, $t_{\rm p}$~=~0.3~d. Using these values, they also reported the expected mass loss rate from prominences for these two stars in absolute units. In order to compare with the steady state wind, we convert their results to units of $\dot{M}_{\odot}$/$R_{\odot}^{\rm 2}$. For DT~Vir we have $\dot{M}_{\bigstar}^{\rm p}$/$R_{\bigstar}^{\rm 2}$~=~0.49~$\dot{M}_{\rm \odot}$/$R_{\odot}^{\rm 2}$ and for GJ~1245~B the resulting value is $\dot{M}_{\bigstar}^{\rm p}$/$R_{\bigstar}^{\rm 2}$~=~0.68~$\dot{M}_{\odot}$/$R_{\odot}^{\rm 2}$.

For the CMEs contribution, we can obtain an order of magnitude estimate by following the approach in \cite{2017MNRAS.472..876O}. They estimate the mass-loss rate from the CME ($\dot{M}_{\bigstar}^{\rm CME}$) as a function of $L_{\rm X}$ and the power law index ($\alpha$) of the flare frequency distribution. For the X-ray luminosity, we used the \cite{NEXXUS2} database, and for the flare frequency distribution exponent we took $\alpha$~=~2 \citep{2014ApJ...797..121H}.

For DT~Vir, with log(L$_{\rm X}$)~=~29.75, we obtain $\dot{M}_{\bigstar}^{\rm CME}$/$R_{\bigstar}^{\rm 2}$~$\sim$~160~$\dot{M}_{\rm \odot}$/$R_{\odot}^{\rm 2}$. For GJ~1245~B, with log(L$_{\rm X}$)~=~27.47, the estimated CME-mass loss rate is $\dot{M}_{\bigstar}^{\rm CME}$/$R_{\bigstar}^{\rm 2}$~$\sim$~12.8$\dot{M}_{\rm \odot}$/$R_{\odot}^{\rm 2}$.

We emphasize here that this approach assumes that the solar flare-CME association rate holds for very active stars (see the discussion in \citealt{2013ApJ...764..170D}). As such, it does not consider the expected influence due to CME magnetic confinement (e.g.~\citealt{2018ApJ...862...93A,2019ApJ...884L..13A}) which currently provides the most suitable framework to understand the observed properties of stellar CME events and candidates \citep{2019ApJ...877..105M,2022AN....34310100A,2022SerAJ.205....1L}.

Still, we can clearly see that the input from CMEs to the total $\dot{M}_{\bigstar}$ could be higher than the steady wind and prominences for these two stars (with the latter contributing less in these cases). For instance, the estimated contribution of CMEs to the total $\dot{M}_{\bigstar}$ of DT~Vir is almost 40 times higher than the value obtained for the steady stellar wind.

We will discuss the cases of EV~Lac and YZ~CMi in Section~\ref{sec:outliers}

\subsubsection{Comparison between simulations and observations}\label{sec:obs-sim}

In addition to analyzing the general trends, we can compare the models for common stars between our sample and the observations in \cite{Wood2021} and references therein. 
The stars in \cite{Wood2021} contain a total number of 37 stars with a mix of main-sequence and evolved stars. The sample includes 15 single K-G stars among them 4 evolved stars, and 4 binaries. \cite{Wood2021} reports individual $\dot{M}$ values for the G-K binary pairs (this means that it was possible to model their individual contribution to the astrosphere of the system or they were separated enough not to share a common astrosphere). This is important as, in principle, one could treat the binary pairs as individual stars. The rest of the star sample includes 22~M-dwarfs with 18 single M-dwarfs, 3~binaries, and 1 triple system. Unlike the G-K stars, $\dot{M}_{\bigstar}$ values for the M-dwarf binaries/triple system are listed as a single value (therefore, it means that it has to be taken as the aggregate of all the stars in the system). For the binary system GJ 338 AB we were unable to include it in the plot of Fig. \ref{fig:fig6} due to a lack of needed information to estimate its $R_{\rm o}$.

Following on the results from Sect.~\ref{sec:Mdot-Ro}, our simulated mass loss rates for stars in the unsaturated regime agree well with those estimated from astrospheric detections (see Fig.~\ref{fig:fig6}). Specifically, for GJ~205~(M1.5V), 61~Cyg~A~(K5V), and HD~219134~(K3V) we obtain $\dot{M}_{\rm \bigstar}/R_{\bigstar}^{\rm 2}$ of 2.32, 3.98, and 1.50, respectively. These values are all consistent with their respective observational estimates, taking into account the typical uncertainties of the astrospheric absorption method\footnote{Astrospheric estimates on $\dot{M}_{\bigstar}$ should have an accuracy of about a factor of 2 with substantial systematic uncertainties \citep{2005ApJ...628L.143W}.}. While further observations could help to confirm this, the agreement between our asynchronous models and the observations indicates that, within this $R_{\rm o}$ range, the temporal variability of $\dot{M}_{\bigstar}$ is minimal. This is certainly the case for the Sun ($R_{\rm o} \sim 2.0$) in which long-term monitoring has revealed only minor variability of the solar wind mass loss rate over the course of the magnetic cycle (\citealt{2011MNRAS.417.2592C}, \citealt{2018ApJ...864..125F,2019ApJ...876...44F}).   

On the other hand, $\dot{M}_{\bigstar}$ from the 3D MHD simulations appear to fall short by an order of magnitude or more from the available estimates for $\epsilon$~Eri~(K2V), EV~Lac~(M3.5V), and YZ~CMi~(M4.5V) with $\dot{M}_{\bigstar}/R_{\bigstar}^{\rm 2}$ of 4.53, 0.62 and 20.57, respectively. We will discuss different possibilities for these discrepancies on each star in Sect.~\ref{sec:outliers}. However, it is important to remember that the $\dot{M}_{\bigstar}$ estimates from the Ly-$\alpha$ absorption technique contain systematic errors that are not easily quantified. One example is that they depend on the assumed properties and topology of the ISM \citep{2014ASTRP...1...43L}, which have not been fully agreed upon in the literature (e.g.,~\citealt{2009ApJ...696.1517K, 2014A&A...567A..58G, 2015ApJ...812..125R}). While studies have provided a detailed characterization of the local ISM (see~\citealt{Redfield_2008, 2015ApJ...812..125R, 2014A&A...567A..58G}), intrinsic uncertainties and additional observational limitations can greatly alter the estimated mass-loss rate values. These include column densities, kinematics, and metal depletion rates (\citealt{2004ApJ...602..776R, 2008ApJ...673..283R}), as well as local temperatures and turbulent velocities \citep{2004ApJ...613.1004R}.

Furthermore, we would also like to emphasize the variation of the  $\dot{M}_{\bigstar}$ in the astrospheric estimates with the assumed stellar wind velocity, as we believe that this factor is one of the largest potential source of uncertainty and discrepancy with our models. As discussed by \citet{Wood2021}, this parameter is used as input in 2.5D hydrodynamic models to quantify the stellar wind mass loss rate. The Ly-$\alpha$ absorption signature, leading to $\dot{M}_{\bigstar}$, is determined to first order by the size of the astrosphere. The latter depends on the stellar wind dynamic pressure ($P_{\rm dyn} \propto \dot{M}_{\bigstar}\,U_{\rm sw}$), which implies an inverse relation between $\dot{M}_{\bigstar}$ and $U_{\rm sw}$ \citep{Wood2002}.

The astrospheric analysis of \citet{Wood2021} assumed a stellar wind velocity of 450 km~s$^{-1}$ at $1$~au (matching models of the heliosphere) for all main-sequence stars. However, we find that stellar wind velocities can vary significantly between different types of stars and even among the same spectral type for different magnetic field strengths and rotation periods. To quantify this, we compute the average terminal velocity of the wind, ($\overline{{U_{\rm R}^{\rm T}}}$), by averaging $U_{\rm R}$ over a sphere extracted at 99$\%$ of the maximum extent of each simulation domain (594~$R_{\rm \bigstar}$ for F, G, and K stars and 248~$R_{\rm \bigstar}$ for M-dwarfs; see Sect.~\ref{Section 2}). In the cases in which the spatial extension of our numerical domain allowed, we also computed the average wind velocity at $1$~au. The resulting values, listed in Table~\ref{tab:2}, indicate variations in the wind velocity by factors of 5 or more when moving from F-type stars ($\overline{U_{\rm R}^{\rm T}} \sim 325$~km~s$^{-1}$) to M-dwarf~($\overline{U_{\rm R}^{\rm T}} \sim 1500$~km~s$^{-1}$). This is also illustrated in Fig.~\ref{fig:fig7}, which portrays the simulated stellar wind environment for HD~179949~(F8V), HD~73256~(G8V), HD~189733~(K2V), and DT~Vir~(M0V). We include a green iso-surface that corresponds to the wind velocity at $1$~au for F, G, and K stars as for M-dwarfs it represents the average terminal wind velocity in the domain. The visualizations also include the equatorial projection of the wind dynamic pressure ($P_{\rm dyn}$~=~$\rho U^{\rm 2}$), normalized to the nominal Sun-Earth value, as well as on a sphere highlighting the wind 3D structure at $0.5$~au.

What is clear from this analysis is that is not ideal to use the same wind velocity for all spectral types. Even within the same spectral type, we can observe a wide range of terminal velocities (e.g., the velocity in K stars ranges from $400$~km~s$^{-1}$ to $700$~km s$^{-1}$). As such, for models that require wind velocity as an input parameter, we recommend using the average radial wind velocity among a given spectral type. 

For G-K stars, we obtain wind velocities at $1$~au in the range of 400 to $700$~km~s$^{-1}$ which is not too different from the wind velocity assumption of \citealt{Wood2021}. This is also consistent with the fact that for these spectral types, we have a better agreement between $\dot{M}$ estimated from our simulations and those from the astropsheric technique \citep{Wood2021}.
For lower mass stars with relatively small $R_{\rm o}$ we obtain velocities higher than $450$~km~s$^{-1}$ up to $3675$~km~s$^{-1}$. 

Note that due to computational limitations, the extent of our M-dwarf simulations does not reach up to $1$~au  (varying from $0.6$~au for DT Vir to $0.16$~au for GJ~1245~B). Nevertheless, as indicated by the calculated terminal velocities, even at closer distances the wind velocity is already $>450$~km~s$^{-1}$, a situation that should still hold when propagated out to $1$~au. Wind velocities on the order of $1000-1500$~km s$^{-1}$ at distances of $1$~au and beyond had been reported in high-resolution AWSoM simulations of the environment around the M5.5V star Proxima Centauri \citep{2020ApJ...902L...9A}. This helps to explain why our simulated mass-loss rates for EV~Lac, YZ~CMi, and $\epsilon$~Eri were lower than the observed ones (differences larger than a factor of 2). We discuss these cases in more detail in the following section.

\subsubsection{Exploring the cases of EV~Lac, YZ~CMi, $\epsilon$~Eri}\label{sec:outliers}

\begin{enumerate}
    
\item \textbf{YZ~CMi \& EV~Lac}

\smallskip
\noindent Frequent stellar flares have been observed at YZ~CMi in several wavelength ranges (\citealt{lacy1976uv, mitra2005relationship, 2013ApJS..207...15K, 2022ApJ...935..102B}). The flaring energy distribution of this star ranges from $10^{30.6}$ to $10^{34.09}$~erg \citep{2022ApJ...935..102B} with a total flaring time that varies from 21 to 306 minutes. Likewise, there is also significant flare activity on EV~Lac (\citealt{leto1997vizier, 2020MNRAS.499.5047M}). From spectroscopic and photometric studies of EV~Lac, \cite{2020MNRAS.499.5047M} reports to have found 27 flares ($\sim$\,5.0~flares per day) in H~$\alpha$ with energies between $1.61 \times 10^{31}$~erg~$−1.37 \times 10^{32}$~erg and 49~flares ($\sim$\,2.6~flares per day) from the TESS lightcurve with energies of $6.32 \times 10^{31}$~erg~$−1.11 \times 10^{33}$~erg.
With such high flare activity, it is possible that a large fraction of the $\dot{M}_{\bigstar}$ estimated in \cite{Wood2021} for these stars could arise from transient phenomena (e.g., prominences, CMEs). 

Following the same approach described at the end of Section~\ref{sec:obs-sim}, we can obtain a rough estimate of $\dot{M}$ from CMEs for EV~Lac and YZ~CMi.
For EV~Lac we find 
$\dot{M}_{\bigstar}^{\rm CME}$/$R_{\bigstar}^{\rm 2}$~=~55.5~$\dot{M}_{\rm \odot}$/$R_{\odot}^{\rm 2}$ assuming log(L$_{\rm X}$)~=~28.69. In the case of YZ~CMi, an log($L_{\rm X}$)~=~28.53 yields $\dot{M}_{\bigstar}^{\rm CME}$/$R_{\bigstar}^{\rm 2}$~=~47.6~$\dot{M}_{\rm \odot}$/$R_{\odot}^{\rm 2}$. However, given the magnetic field strength observed in EV~Lac and YZ~CMi (a few kG, \citealt{2010A&A...523A..37S,2022A&A...662A..41R,2023MNRAS.522.1342C}), we expect that the magnetic confinement of CMEs would play an important role in these objects (see \citealt{2019ApJ...884L..13A}, \citealt{2020MNRAS.499.5047M}, \citealt{2021PASJ...73...44M}). Therefore, it is not straightforward to estimate exactly how large the contribution of CMEs to $\dot{M}_{\bigstar}$ is for these stars.

In addition, as discussed by \citet{Villarreal2018}, EV~Lac and YZ~CMi are considered in the slingshot prominence regime. For EV~Lac they estimate $m_{\rm p} =  2.0 \times 10^{\rm 16}$~g and $t_{\rm p}$~=~0.6~d, while for YZ~CMi values of $m_{\rm p} =  4.5 \times 10^{\rm 16}$~g and $t_{\rm p}$~=~0.6~d are given. Using the associated mass loss rate values reported in \cite{Villarreal2018}, we obtain $\dot{M}_{\bigstar}^{\rm p}$/$R_{\bigstar}^{\rm 2}$~=~3.16~$\dot{M}_{\rm \odot}$/$R_{\odot}^{\rm 2}$ for EV~Lac and $\dot{M}_{\bigstar}^{\rm p}$/$R_{\bigstar}^{\rm 2}$~=~8.32~$\dot{M}_{\rm \odot}$/$R_{\odot}^{\rm 2}$ for YZ~CMi.

This suggests another possible explanation for the discrepancies between our models and the astrospheric estimates is that some of the stellar wind detected for EV~Lac and YZ~CMi contains material from the slingshot prominences. Indeed, the location of the latter in the $\dot{M}_{\bigstar}$~--~$R_{\rm o}$ diagram (Fig.~\ref{fig:fig6}) appears more consistent with the mass loss rate estimates from slingshot prominences by \citet{2019MNRAS.482.2853J}. 

Moreover, \cite{Wood2021} noted that the YZ~CMi astrospheric absorption comes primarily from neutrals near and inside the astropause, rather than from the hydrogen wall where neutral H density is highest. Therefore, using Ly alpha absorption to calculate $\dot{M}_{\bigstar}$ from YZ~CMi will result in substantial uncertainty.

Finally, as mentioned in Sect.~\ref{sec:obs-sim}, there is a significant difference between the wind velocity assumed by \cite{Wood2021} and our results. Our average terminal wind velocity for YZ~CMi ($1709$~km~s$^{-1}$) and EV~Lac ($3675$~km~s$^{-1}$) is significantly higher than the wind velocity of $450$~km~s$^{-1}$ assumed in \cite{Wood2021} at 1 au. While the wind velocity in EV~Lac might be overestimated in our models (due to the usage of fiducial AWSoM parameters), we still expect relatively large wind velocities for this star ($\sim$\,$1000-1500$~km~s$^{-1}$) given its magnetic field strength and Rossby number (see e.g.,~\citealt{2021MNRAS.504.1511K,2022ApJ...928..147A}). As was discussed in Sect.~\ref{sec:obs-sim}, while our terminal wind velocity for M-dwarfs is calculated closer to the star ($0.33$~au for YZ~CMi and $0.16$~au for EV~Lac), we do not expect a large reduction in the average velocity between these distances and 1 au. As such, the fast wind velocity resulting in our simulations of YZ~CMi and EV~Lac would imply lower $\dot{M}_{\bigstar}$ values when analyzed following the astrospheric technique of \cite{Wood2021}.

\smallskip

\item \textbf{$\epsilon$~Eri}

\smallskip

\begin{figure*}
\includegraphics[width=1\textwidth]{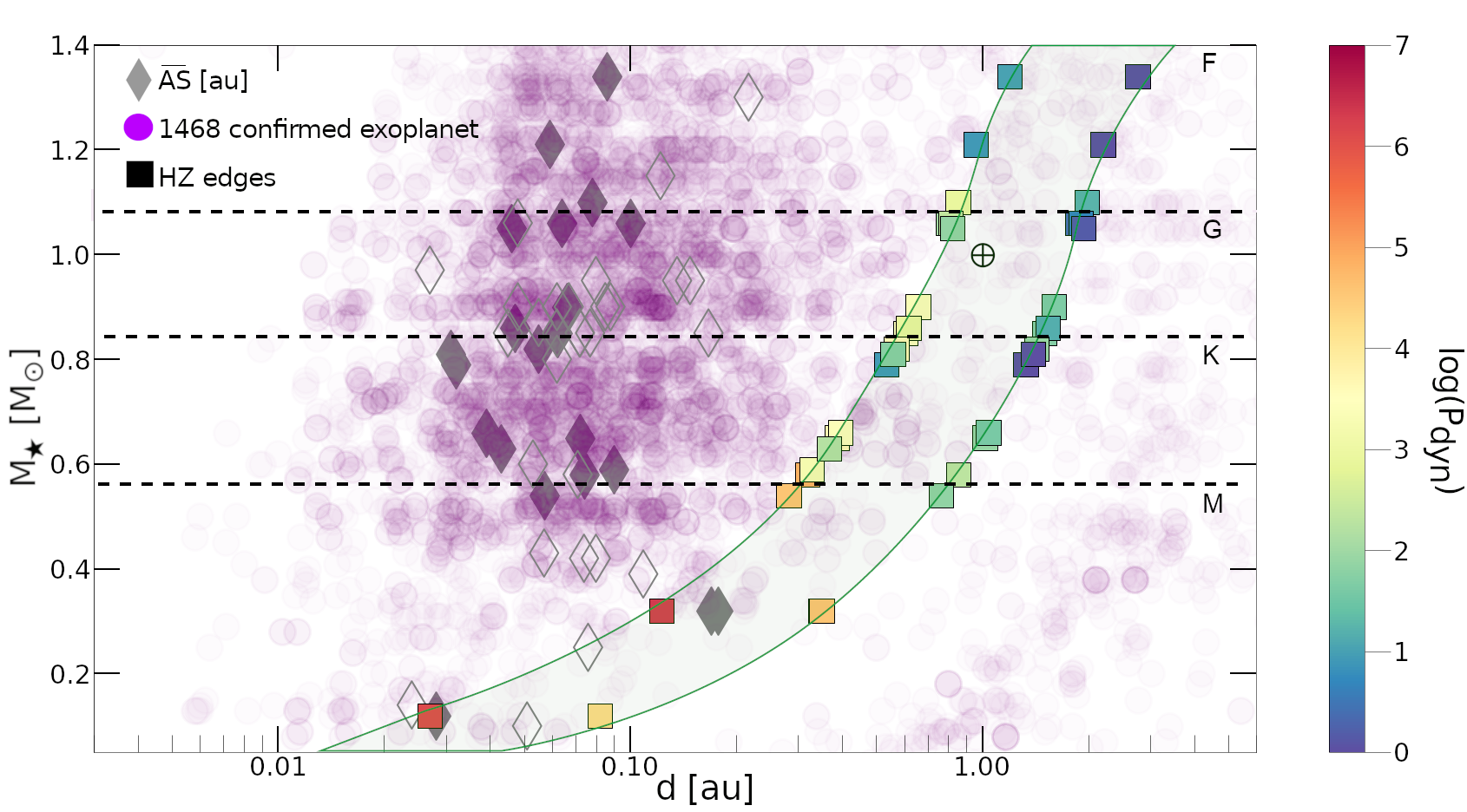}
    \caption{ Numerical results of the average Alfv\'{e}n surface size (diamonds), the inner and outer edge of the HZ (square, \citealt{Kopparapu2014}) derived using scaling laws that connect $M_{\bigstar}$, $R_{\bigstar}$, $L_{\bigstar}$, and $T_{\rm eff}$, against $M_{\bigstar}$. The filled diamonds  correspond to the $\overline{AS}$ of the 21 stars in our sample. The empty diamonds are the $\overline{AS}$ of other stars derived from our $\overline{AS}$ - $B_{\rm R}$ relation: log $\overline{AS_{\rm R}} = (0.42 \pm 0.06) \log B_{\rm R}^{\rm avg} + (0.71 \pm 0.07)$. The habitable zone boundaries are color-coded by the corresponding average dynamic pressure ($\overline{P_{\rm dyn}}$) in logarithmic scale. The green shaded area represents the optimistic habitable zone. Earth is represented by $\oplus$. The purple circles correspond to some confirmed exoplanets (taken from the NASA exoplanet archive). The dynamic pressure at the outer edge of the HZ of DT~Vir and GJ~205~(derived from scaling laws) is missing in the plot since it goes beyond the simulation domain that was initially established using the measured $L_{\bigstar}$ and $T_{\rm eff}$ (Table~\ref{tab:1}). The dashed lines separate the different spectral types.}
    \label{fig:fig8}
\end{figure*} 

\noindent With a relatively slow rotation period ($11$~d), and weak large-scale magnetic field ($<$\,50~G), $\epsilon$ Eri cannot be considered within the slingshot prominence regime (like in the cases of YZ~CMi and EV~Lac). Because of this, we do not expect a significant presence of slingshot prominences in the $\dot{M}$ value of this star. On the other hand, the analysis of \citet{2022ApJ...936..170L}, estimated the contribution of flare-associated CMEs to the mass loss rate. They reported an upper limit of $1.09 $~$\dot {M}_{\rm \odot}$~$/ R _{\odot}^{\rm 2}$, which is insignificant when compared to the star's overall estimated $\dot {M}_{\bigstar}$ value by \cite {Wood2021} and the astrospheric technique (56~$\dot {M}_{\rm \odot}$~$/ R _{\odot}^{\rm 2}$).
Therefore, the contribution from CMEs is also most likely not responsible for the elevated astrospheric $\dot{M}_{\bigstar}$ value on this star and its discrepancy with our steady-state models.

On the other hand, multiple observations of the large-scale magnetic field geometry of $\epsilon$~Eri reveal that it evolves over a time-scale of months \citep{2014A&A...569A..79J, Jeffers2017}. According to \cite{2014A&A...569A..79J}, the maximum field strength can reach up to $42$~G. As shown in Fig.~\ref{figure3}, a global increase in the magnetic field strength causes an increase in $\dot{M}_{\bigstar}$. 
The Zeeman Doppler Imaging map of $\epsilon$~Eri used to drive the 3D MHD model has a $B_{\rm R}^{\rm max}$~=~25~G leading to $\dot{M}_{\bigstar}/R_{\bigstar}^{2} = 4.53~\dot{M}_{\rm \odot}/R_{\rm \odot}^{2}$. This value is comparable to the numerical result obtained by  \cite{2016A&A...594A..95A} for this star ($\dot{M}_{\bigstar}/R_{\bigstar}^{2} \sim 5.3 \dot{M}_{\rm \odot}/ R_{\rm \odot}^{\rm 2}$). Increasing the surface magnetic field strength of $\epsilon$~Eri to the maximum value reported in observations will raise the mass loss rate to $\sim 10~\dot{ M}_{\rm \odot}/ R_{\rm \odot}^{\rm 2}$. As such, the variability of the stellar magnetic field and its expected modulation of the stellar wind properties could account for some of the differences between the simulated and the observed mass loss rates. However, corroborating this would require contemporaneous ZDI and astrospheric measurements which, to our knowledge, have not been performed on any star so far.
As $\epsilon$~Eri goes through a magnetic/activity cycle (\citealt{2013ApJ...763L..26M, Jeffers2017}), we can expect relatively large variations in $\dot{M}_{\bigstar}$ values in our Alfv\'{e}n-wave driven stellar wind models.

Finally and following the discussion for YZ~CMi and EV~Lac, the average wind velocity for $\epsilon$~Eri at $1$~au ($554$~km~s$^{-1}$) resulting from our models exceeds the one assumed in \cite{Wood2021}. This will result in a smaller estimated $\dot{M}_{\bigstar}$ value from the pressure-balance astrospheric technique. 
In this way, the deviation between our models and the astrospheric detection of $\epsilon$~Eri could be due to the combined contribution of all the preceding elements (i.e.,~CMEs, cycle-related variability of the magnetic field, higher stellar wind velocity), and therefore we do not consider this discrepancy critical to our analysis.

\end{enumerate}

\subsection{Stellar wind and Circumstellar region} 

This section focuses on using the stellar wind results obtained from the 3D MHD simulations to assess the conditions an exoplanet would experience. This includes the characterization of the Alfv\'{e}n surface for the various stellar wind solutions, the properties of the stellar wind in the habitable zone of these stars (in terms of the dynamical pressure of the wind), and the resulting magnetosphere size for these stellar wind conditions (assuming that a planet with the same properties/magnetization as Earth is in the HZ of these stars). The obtained quantities are listed in Table~\ref{tab:2} and \ref{table3}. 

\subsubsection{Stellar wind properties and orbital distances}

\begin{enumerate}
    \item \textbf{Alfv\'{e}n surface size}

\noindent Figure~\ref{fig:fig8} summarizes our results showing the stellar wind environment around cool main sequence stars. We include the average size of the AS, resulting from our 21 3D MHD models, indicated in filled diamonds. To complement this information, empty diamonds correspond to the expected average AS size employing the scaling relation provided in Sect.~\ref{sec:3.1}, and using the ZDI information from 29 additional stars (\citealt{See2019_magneticfield} and reference therein). The green region corresponds to the optimistic HZ, calculated using the approach provided by \cite{Kopparapu2014} and the expected behaviour of the luminosity, temperature as a function of stellar mass on the main sequence (\citealt{KASTING1993108,Kopparapu2014,2018Geosc...8..280R}). Each square indicates the limits of the optimistic HZ for each star in our sample. These have been color-coded by the stellar wind dynamic pressure, normalized to the average Sun-Earth value. The position of the Earth is indicated by the $\oplus$ symbol. In the background, a sample of the semi-major axis of some exoplanets is included.

There are a few noteworthy aspects of Fig.~\ref{fig:fig8}. First of all, the 3D MHD simulated $\overline{AS}$ values (filled diamonds) do not show a clear trend with stellar mass. Instead, we see more or less similar $\overline{AS}$ regardless of the spectral type of the star (Table~\ref{tab:2}). We see a similar behavior for stars whose $\overline{AS}$ were extracted from the scaling relationship  presented in Eq.~\ref{eq3}~(empty diamonds). There is a significant scatter in the obtained distribution of $\overline{AS}$ against $M_{\bigstar}$, indicating that the intrinsic dependency with the surface magnetic field properties can in principle be replicated among multiple spectral types. However, we remind the reader that this result is also partly a consequence of our fixed choice for the base parameters of the corona and stellar wind solution (Sect.~\ref{sec:iParams}), which could in principle vary among different spectral types and activity stages (i.e. ages). As such, the generalization of the results presented here requires further investigation from both, observational constraints and numerical simulations.

We can also see that for late K and M-dwarfs, $\overline{AS}$ reaches orbital distances comparable to their HZ limits. Examples of this from our sample are GJ~1245~B~($\overline{AS}$~=~0.028~au, $HZ_{\rm inner}$~=~0.033~au) and YZ~CMi ($\overline{AS}$=~0.178~au, $HZ_{\rm inner}$=~0.09~au). This situation has been also identified in previous case studies of stellar winds and exoplanets (e.g,~\citealt{2013A&A...557A..67V,2014ApJ...790...57C,2017ApJ...843L..33G}). 

The location of the HZ relative to the stellar Alfv\'{e}n surface must be considered when studying the interactions between a star and a planet. A planet orbiting periodically or continuously within the $AS$ region could be directly magnetically connected to the stellar corona, which could have catastrophic effects on atmospheric conservation \citep{2014ApJ...790...57C,2017ApJ...843L..33G, 2021arXiv210405968S}.
On the other hand, a planet with an orbit far outside this limit will be decoupled from the coronal magnetic field and interact with the stellar wind in a manner similar to the Earth (e.g. \citealt{2020ApJ...897..101C, 2020ApJ...902L...9A}). In the case of a planet orbiting in and out of the AS, the planet will experience strongly varying wind conditions, whose magnetospheric/atmospheric influence will be greatly mediated by the typical time-scale of the transition \citep{2021ApJ...913..130H}. 

\begin{table*}
\centering
\caption{Numerical results of different parameters in our sample at the habitable zone. Columns 1–8, respectively, list the star name, average dynamic pressure at the middle of the HZ~($\overline{P}_{\rm dyn, HZ}$), average dynamic pressure at the inner boundary of the HZ~($\overline{P_{\rm dyn}^{\rm Inn, HZ}}$), average dynamic pressure at the outer boundary of the \textit{HZ }($\overline{P_{\rm dyn}^{\rm Out, HZ}}$), the average magnetopause standoff radius ($\overline{ R}_{\rm M}$), the inner habitable zone~($HZ_{\rm inner}$), the outer habitable zone~($HZ_{\rm outer}$), the average equatorial Alfv\'{e}n surface ($\overline{AS}_{\rm eq}$). The habitable zones listed in this table were inferred using the measured $L_{\bigstar}$ and $T_{\rm eff}$ of each star in our sample (Table~\ref{tab:1}). }

\label{table3}
\begingroup

\setlength{\tabcolsep}{10pt} % Default value: 6pt
\renewcommand{\arraystretch}{1.5} % Default value: 1
\resizebox{0.9\textwidth}{!}{%
\begin{tabular}{p{23mm}|p{10mm}|p{10mm}||p{10mm}|p{10mm}|p{10mm}|p{10mm}|p{5mm}|p{5mm}}
\hline
Name &  ${\overline{P}_{\rm {dyn, HZ}}}$ [$ P_{\rm \oplus}$] & $\overline{P_{\rm dyn}^{\rm Inn, HZ}}$ [$P_{\rm \oplus}$]& $\overline{P_{\rm dyn}^{\rm Out, HZ}}$ [$P_{\rm \oplus}$]&$\overline{
R}_{\rm M, HZ}$ [$R_{\rm E}$] &  $HZ_{\rm inner}$ [au] & $HZ_{\rm outer}$ [au]& $  \overline{AS}_{\rm eq}$ [$R_{\rm \bigstar}$]\\
\hline % inserts single horizontal[3ex] \midrule 

$\tau$~Boo   &1.02  & 2.76 & 0.40 &7.85 & 1.253 & 2.901  & 9.99\\

HD~179949    &0.92 & 2.51& 0.48 &7.99& 0.983 & 2.294 & 10.24 \\

HD~35296     & 2.93  & 7.77& 1.48 &6.58 & 0.925 & 2.157& 12.78 \\

HN~Peg       & 6.74 & 18.58 & 3.47  & 5.73 & 0.812  &  1.904 & 13.65 \\

HD~190771    & 3.96 & 11.01& 2.02  & 6.26 & 0.744 &  1.748 &  12.35\\

TYC~1987-509-1 & 10.21 & 28.49 & 5.20 &5.30& 5.35  & 1.299 & 15.47  \\

HD~73256     & 2.32 & 6.45 & 1.18 & 6.85& 0.648 & 1.539 & 10.73  \\

HD~130322      & 0.91 & 2.56 &  0.46 &8.00 & 0.542  & 1.291  & 8.12 \\

HD~6569        & 10.77 & 30.21  &5.46 &5.30 &0.466 & 1.11 & 16.25  \\

$\epsilon$~Eri  &6.45 & 18.16 &  3.23  & 5.77& 0.426 & 1.024 & 13.64  \\

HD~189733       & 13.56& 38.82 &  6.88 & 5.10& 0.458& 1.108  & 13.4  \\

HD~219134    & 2.01 & 5.81  &1.01 & 7.01&  0.410  & 0.996 & 8.33 \\

TYC 6878-0195-1 & 13.42&  39.39 &6.71 & 5.11&0.713 &1.751 & 23.16  \\

61~Cyg~A        & 10.06 & 29.47   & 5.01  & 5.36& 0.308 &  0.754 & 11.93  \\

HIP 12545       & 20.28&  60.51 &  9.98 & 4.77& 0.509 &1.267  & 24.45 \\

TYC 6349-0200-1 & 13 & 40 & 6 & 5.14& 0.443&1.116  & 21 \\

DT Vir  & 31.48  & 88.97  &  16.20 & 4.43&  0.191 & 0.486 & 29.24  \\

GJ~205   & 9.73& 29.65  &4.73  & 5.39& 0.201&  0.517 & 15.91  \\

EV~Lac  &  12.15& 33.18  & 6.27& ---- & 0.094& 0.245 & 119.16 \\

YZ~CMi  &193.85 & 566.34  & 97.08 &---- & 0.09 & 0.237& 75.23  \\

GJ~1245~B  & 140.89 & 447.39 & 68.15 &---- & 0.033 & 0.087 & 38.12\\
\hline
\end{tabular}%

}
\endgroup
\end{table*}

\begin{figure}% <---
           % <--- defined in "sidecap" package
     \includegraphics[width=1\columnwidth]{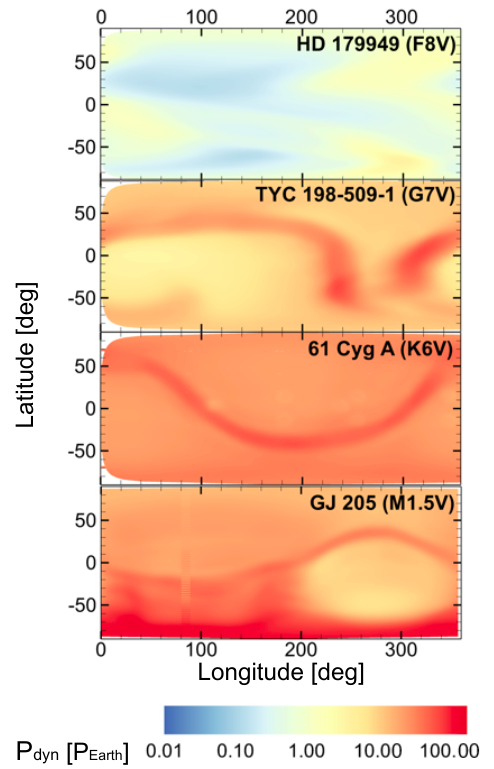}
    \caption{Two-dimensional Mercator projections of the normalized stellar wind dynamic pressure ($P_{\rm dyn}$) extracted from the 3D MHD models of four stars in our sample covering F to M spectral types~(HD~179949, TYC~198-509-1, 61~Cyg~A, and GJ~205). Each $P_{\rm dyn}$ distribution was extracted from a spherical surface located at the midpoint habitable zone of each star. }
    \label{fig:fig9}
\end{figure}

 \item \textbf{Dynamic pressure }

\noindent We also see a general trend in Fig.~\ref{fig:fig8} in which the dynamic pressure at the HZ boundaries increases as we move from earlier to later spectral types. For example, $\overline{P_{\rm dyn}^{\rm Inn, HZ}}$ for the lowest-mass star GJ~1245~B is 447.39~$P_{\rm \oplus}$ nearly 200 times stronger than for the highest-mass star $\tau$~Boo with $\overline{P_{\rm dyn}^{\rm Inn, HZ}}$~=~2.76~$P_{\rm \oplus}$.
Our results also show a large variability in $\overline{P_{\rm dyn}}$ as we move from the inner to the outer edge of the HZ of G, K, and M-dwarfs (Table~\ref{table3}). For these stars, the $\overline{P{\rm dyn}}$ at the inner HZ is almost 6 times stronger than that at the outer edge of the HZ (i.e.,~EV~Lac $\overline{P_{\rm dyn}^{\rm Inn, HZ}}$~=~33.18~$P_{\rm \oplus}$, $\overline{P_{\rm dyn}^{\rm Out, HZ}}$~=~6.27~$P_{\rm \oplus}$). For F stars, the difference is smaller, around a factor of 2 like in the case of HD~179949, where $\overline{P_{\rm dyn}^{\rm Inn, HZ}}$~=~2.51~$P_{\rm \oplus}$ and $\overline{P_{\rm dyn}^{\rm Out, HZ}}$~=~0.48~$P_{\rm \oplus}$. The reason is that the HZs of these stars are farther from the star, where the wind density starts to become less variable.

Moreover, in some cases, we have $\overline{P}_{\rm dyn}$ at the inner and outer edge of the star HZ comparable to the typical range experienced by the Earth (0.75 and 7 nPa, \citealt{2021SSRv..217...36R}). For example, HD~73256~(G8V,~6.45~-~1.18~$P_{\rm \oplus}$~$\sim$~9.675~-~1.77~nPa), HD~130322 (K0V,~2.56~-~0.46~$P_{\rm \oplus}$~$\sim$~3.84~-~0.69~nPa), $\tau$Boo (F7V,~2.76~-~0.40~$P_{\rm \oplus}$~$\sim$~4.14~-~0.6~nPa). For the case of M-dwarfs, we have dynamic pressures higher than those experienced by Earth, as in the case of DT Vir (M0V,~88.97~-~16.20~$P_{\rm \oplus}$~$\sim$~133.455~-~24.3~nPa). This is because the HZ is located near the star where the density is highest. This indicates that planets orbiting at very close distance to the star ($\sim$~0.03~-~0.05~au) would experience extreme space weather conditions with ${P}_{\rm dyn}$ up to $\rm 10^{3}$ and $\rm 10^{4}$ $P_{\rm \oplus}$. These values are comparable to the ones estimated in \cite{2020ApJ...902L...9A} for Proxima d and for Proxima b in \cite{2016ApJ...833L...4G}. However, the reader is reminded here that any point from our simulations should be interpreted as an indication of the average conditions, but should not be treated as a specific absolute value (since it will change depending on the instantaneous local density and velocity of the wind (both a function of the evolving stellar magnetic field).

In addition, we notice a scatter in $\overline{P}_{\rm dyn}$ estimates at the HZ when comparing stars of the same spectral type. This is not surprising since the $P_{\rm dyn}$ depends on the wind velocity and density at a given place. This also translates into having a range of dynamic pressure that a planet will experience within the HZ. This will defer from one orbital distance to the other as we can see in Fig. \ref{fig:fig7} where we show the equatorial plane color-coded by the dynamic pressure.

We can use our 3D models to investigate also the influence due to the orbital inclination. To illustrate this, Fig.~\ref{fig:fig9} shows a 2D projection of the normalized dynamic pressure $P_{\rm dyn}$ extracted from spherical surfaces matching the midpoint of the HZ of HD~179949~(F8V), TYC~198-509-1~(G7V), 61~Cyg~A~(K6V), and GJ~205~(M1.5V). We notice that in the case of F and G stars (i.e.,~HD~179949, and TYC-198-509-1) we have a large $P_{\rm dyn}$ variation with inclination around a factor 7. However, $P_{\rm dyn}$ values, are still relatively small in terms of absolute units (i.e.,~0.01~-~10~$P_{\rm \oplus}$ $\sim$~0.015~-~15~nPa). For K and M-dwarfs, we see less variability in the $P_{\rm dyn}$ for the different inclinations, a more homogeneous $P_{\rm dyn}$, especially in the case of the K star. However, in these cases, the $P_{\rm dyn}$ can reach values >~100~$P_{\rm \oplus}$~(>~150~nPa). Our results also show that even with an extreme orbit around the G-type star (TYC~198-509-1) with an inclination matching the current sheet, we would most likely not reach the very high $P_{\rm dyn}$ values as in the case of the K and M-dwarfs as we move closer to the star. As such, the inclination of the orbit plays a secondary role compared to the distance. This is clearly seen in the color gradient that gets redder and redder as we move toward lower masses (so the HZ is closer).

On the other hand, the variability of $P_{\rm dyn}$, which we can see in Fig.~\ref{fig:fig8} while represented in the same 'spatial scale', it does not coincide in terms of 'temporal scales'. In other words, the $x$-axis in Fig.~\ref{fig:fig8} do not correspond to the same timescale units for each star, where the 360 degrees of longitude correspond to ``1 orbital period''. However, the orbital period is very different for a planet in the HZ of an F-type star (within a few au) compared to a planet orbiting an M-dwarf (within a fraction of an au). A planet orbiting an M-dwarf star experiences the variations in $P_{\rm dyn}$ on a much faster timescale ($\sim$ 1 day for each current sheet crossing), while these variations are much longer for more massive stars. This means that even if the $P_{\rm dyn}$ values were the same, the faster variability over the orbital period for low-mass stars would result in planets and their magnetospheres/atmospheres having less time to recover from passing through regions of high $P_{\rm dyn}$ than planets around more massive stars.

Finally, following the results compiled by \cite{2021SSRv..217...36R}, if we consider the presence of a rocky exoplanet with an atmosphere similar to those of Venus and Mars at those mid-HZ locations, we would expect atmospheric ion losses between $2 \times 10^{24}$~ions~s$^{-1}$ and $5 \times 10^{24}$~ions~s$^{-1}$. This of course assumes that all processes occur in the same way as in the solar system (which might not be necessarily true for some regions of the vast parameter space of this problem). The ion losses will depend heavily on the type of stars that the exoplanet orbits, both in terms of the high-energy spectra and the properties of the stellar wind (see e.g.~\citealt{2019MNRAS.486.1283E, 2020AJ....160..237F}). If the rocky exoplanet is found around the HZ of an M-dwarf, the planet might suffer from unstable stellar wind conditions as previously stated that might increase the ion losses in the exoplanetary atmosphere. We will consider the case of Earth with its magnetosphere in the following section.

\textcolor{teal}{}
\end{enumerate}
\subsubsection{Magnetopause Standoff Distances}

Using the dynamic pressure, we can define a first-order approximation to determine the magnetosphere standoff distance ($R_{\rm M}$) of a hypothetical Earth-like planet orbiting at the HZ around each star in our sample. This is done by considering the balance between the stellar wind dynamic pressure and the planetary magnetic pressure (Eq.~\ref{eq8}, \citealt{2004pse..book.....G,2016PhR...663....1S}):

\begin{equation}
    R_{\rm  M} = R_{\rm E} [\frac{{B_{\rm p}}^{2}}{ 8 \pi P_{\rm dyn}}]^{\frac{1}{6}} \label{eq8}
\end{equation}

\noindent The Earth's equatorial dipole field and radius are represented by $B_{\rm p}$ and $R_{\rm E}$ respectively. Normally the total wind pressure should be considered (i.e.,~thermal, dynamic, and magnetic), but in all the cases here considered, we can neglect the contributions of the magnetic and thermal pressures. For this calculation, we assume an equatorial dipole magnetic field of 0.3~G, similar to that of the Earth \citep{2007LRSP....4....1P}.
The magnetospheric standoff distance is expressed in Earth's radii (Eq.~\ref{eq8}). The different $\overline{R}_{\rm \scriptscriptstyle M, HZ}$ values for the different stars in our sample are listed in table \ref{table3}. Note that we only estimate the $R_{\rm M}$ in the cases where the HZ is in the super-Alfv\'{e}nic regime (\citealt{2014ApJ...790...57C, 2018haex.bookE..25S}).

Our estimated $R_{\rm M, HZ}$ for F, G, and early K stars have values closer to the standard size of Earth's dayside magnetosphere ($\sim$~10~$R_{\rm \oplus}$, see \citealt{2007LRSP....4....1P, Lugaz2015}). This is comparable to the value obtained by \cite{2020ApJ...902L...9A} for Proxima c ($\sim$ 6~-~8~$R_{\rm \oplus}$ in both activity levels), assuming an Earth-like dipole field on the planet surface.
For the late K and M-dwarfs in our star sample, $R_{\rm M}$ starts to reach lower values <~50$\%$ from that of Earth.
This suggests that a planet orbiting these stars must have a stronger dipole magnetic field than that of the Earth to withstand the wind conditions since $R_{\rm M} \propto B_{\rm p} ^{1/3}$. 
However, in \cite{2021SSRv..217...36R} they show that contrary to what we have seen so far, the magnetosphere might actually not act as a shield for the stellar wind-driven escape of planetary atmospheres. In fact, they reported an ion loss for Earth that ranges from 6~$\times 10^{24}$~ions~s$^{-1}$~-~6~$\times 10^{26}$~ions~s$^{-1}$ which is higher than what Venus and Mars lose. Further modeling studies are needed in order to characterize the stellar wind influence on the atmospheric loss of rocky exoplanets (e.g.,~\citealt{Egan2019, 2020AJ....160..237F}), whose input stellar wind parameters can be extracted from this investigation.   

\section{Summary \& Conclusions}
\label{Section 4}

In this study we employed a state-of-the-art 3D MHD model (SWMF/AWSoM) to investigate the dependencies between different star properties ($R_{\bigstar}$, $M_{\bigstar}$, $B_{\rm R}$, and $P_{\rm rot}$) and a number of stellar wind parameters (AS, $\dot{M}_{\bigstar}$, $\dot{J}_{\bigstar}$, $P_{\rm dyn}$) of cool main sequence stars. We present numerical results of 21 stars going from F to M stars with magnetic field strengths between 5 and 1.5~kG and rotation periods between 0.71~d and 42.2~d.
The large-scale magnetic field distribution of these stars, obtained by previous ZDI studies, were used to drive the solutions in the Stellar Corona domain, which are then self-consistently coupled for a combined solution in the Inner Astrosphere domain in the case of F, G, and K stars.

Our results showed a correlation between the average AS size and $B_{\rm R}^{\rm avg}$, regardless of the spectral type of the star (Eq.~\ref{eq3}). We also obtained a strong correlation between $\dot M_{\bigstar}$ and $B_{\rm R}^{\rm avg}$ for the different spectral types (excluding EV~Lac, Eq.~\ref{equ4}). The correlation between $\dot{J}_{\bigstar}$ and $B_{\rm R}$, on the other hand, was dominated by the absolute dependence on the stellar size, with significant scatter resulting mainly from the variability in $\dot{M}_{\bigstar}$, the distribution of $\Omega_{\bigstar}$ in our sample and the equatorial AS size where the maximum torque is applied.

Having established these star-wind relations, we looked in detail at $\dot{M}_{\bigstar}$, since it is the only observable parameter of the stellar wind for which comparisons can be made. Using the complexity number as a function of the Rossby number $R_{\rm o}$--defined previously in the literature-- we were able to investigate the dependence of magnetic complexity on $\dot{M}_{\bigstar}$. Our results showed that for more active stars, as in the case of M-dwarfs, the field strength starts to dominate over the complexity in the contribution on shaping $\dot{M}_{\bigstar}$. Also, for cases in which the magnetic field strength and complexity were comparable, we obtained similar $\dot{M}_{\bigstar}$. This indicates that in these cases the stellar properties ($R_{\bigstar}$, $M_{\bigstar}$, and $P_{\rm rot}$) play a secondary role in changing $\dot{M}_{\bigstar}$.

We then used our stellar wind results to investigate its behaviour with respect to the well-known stellar activity relationship ($F_{\rm X}$~vs~$R_{\rm o}$ with the saturated and unsaturated regimes). For stars in the unsaturated regime, we see a trend where $\dot{M}_{\bigstar}$ increases with decreasing $R_{\rm o}$ (Eq.~\ref{eq7}). For stars in the saturated regime, we find that the contribution of the steady wind is only a small part of the $\dot{M}_{\bigstar}$ budget. This suggests that there could be saturation in $\dot{M}_{\bigstar}$ due to the steady stellar wind, while the star could lose even more mass through other mechanisms, such as transient events (i.e. prominences, coronal mass ejections). 

In addition to analyzing the general trends, we compared the model results of stars in our sample and objects with astrospheric $\dot{M}_{\bigstar}$ constraints. Our simulated $\dot{M}_{\bigstar}$ for stars in the unsaturated regime agree well with those estimated from astrospheric detections (namely for GJ~205, 61~Cyg~A, and HD~219134). On the other hand, $\dot{M}_{\bigstar}$ from the 3D~MHD simulations appear to differ by an order of magnitude or more from available estimates for $\epsilon$~Eri, EV~Lac, and YZ~CMi. We discussed how these results might be connected with the underlying assumption made by the observational analysis with respect to the stellar wind speed. Indeed, for all the stars in which our models differed largely from the literature estimates, we obtained much larger stellar wind speeds than the ones used in the astrospheric method. As such, we emphasized the importance of using the appropriate wind velocity when estimating $\dot M_{\bigstar}$ from observations. 

We further discussed various possibilities for the discrepancies in EV~Lac, YZ~Cmi, $\epsilon$~Eri. For the two flaring stars, EV~Lac and YZ~CMi, we suspect that the high $\dot{M}_{\bigstar}$ estimates from the Ly-$\alpha$ absorption technique could be dominated by material from slingshot prominences and possibly CMEs (uncertain due to the expected magnetic confinement of CMEs in these stars). Note that this possibility was also considered by \cite{Wood2021} in the original astrospheric analysis. In the case of $\epsilon$~Eri, we do not expect a large contribution from prominences or CMEs to the observed $\dot M_{\bigstar}$. However, as $\epsilon$~Eri undergoes a magnetic cycle, the stellar magnetic field and its expected modulation of stellar wind properties could explain some of the differences between the simulated and observed $\dot M_{\bigstar}$. 

Moreover, we used the stellar wind results from the 3D MHD simulations to assess the conditions that an exoplanet would experience, and provide the stellar wind conditions in the entire classical Habitable Zones of our target stars. Our results show a scatter in the obtained distribution of AS versus $M_{\bigstar}$, suggesting that the intrinsic dependence with the surface magnetic field properties can be reproduced for several spectral types. With respect to the stellar wind dynamic pressure, our results show that the orbital inclination plays a secondary role compared to the orbital distance. We have also found that a planet orbiting K and M stars must have a stronger dipole magnetic field than that of Earth to withstand the wind conditions, if the planetary magnetic field is indeed acting as a shield (this paradigm, however, is starting to be challenged by solar system observations).

Finally, the properties of the stellar wind in the HZ of different spectral types obtained here can \textcolor{red}{be} used in future studies to, for instance, estimate the expected radio emission due to wind-magnetosphere interactions or the planetary atmospheric mass loss due to erosion of the stellar wind from ion escape processes.

\section*{Acknowledgements}
The authors would like to thank the referee for valuable comments that improved the quality of the paper.
The authors gratefully acknowledge the Gauss Centre for Supercomputing e.V. (\url{www.gauss-centre.eu}) for funding this project by providing computing time on the GCS Supercomputer SuperMUC-NG at Leibniz Supercomputing Centre (\url{www.lrz.de}) under application ID 21761 (PI: Alvarado-G\'omez). JJC and KP acknowledge funding from the German \textit{Leibniz Community} under project number P67/2018. CG was supported by NASA
contract NAS8-03060 to the Chandra X-ray Center. This research has made use of NASA’s Astrophysics Data System Bibliographic Services.

\section{Data availability}
The data would be made available to the community on reasonable request due to the volume of the 3D simulations. Extractions of specific quantities discussed in the paper could be requested from the corresponding author.


\bibliographystyle{mnras}
\bibliography{example} % if your bibtex file is called example.bib

\appendix
\section{Trends with Maximum Radial Magnetic Field}
\label{sec:appendix}

\begin{figure}
 \centering
         \includegraphics[width=1\columnwidth]{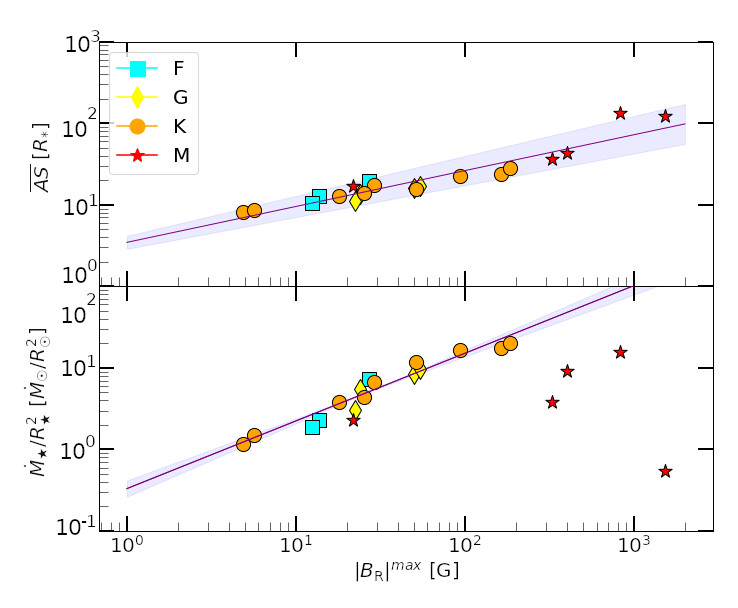}
         \caption{Simulated average Alfv\'en surface ($\overline {AS}$, top), and the mass loss rate per unit surface area ($\dot{M}_{\bigstar}/R^{\rm 2}_{\rm \bigstar}$, bottom) as a function of the absolute surface-max radial magnetic field ($|B_{\rm R}|^{\rm max}$). The mass loss rate is expressed in units of the average solar values, normalized to the surface area of each star. Individual points denote the results of each simulation presented in Sect. \ref{Section 3}, Table \ref{tab:2}. The different symbols and colors represent the spectral types (F, cyan/squares; G, yellow/diamonds; K, orange/circles; M, red/star). The purple line with the shaded purple area represents the fitted power-law with its uncertainties. \label{fig10}}
\end{figure}

We have also quantified $\overline{ AS}_{\rm R}$ and $\dot{M}_{\rm \bigstar}$ / $R^{\rm 2}_{\bigstar}$ as a function of the absolute maximum radial magnetic field strength ($|B_{\rm R}|^{\rm max}|$). It is important to also investigate $|B_{\rm R}|^{\rm max}$, since the average radial magnetic field strength may suffer from cancellations, especially if the star has a symmetric surface magnetic field distribution. Figure \ref{fig10} shows the simulated average Alfv\'en surface area ($\overline {AS}$, top) and the mass-loss rate per unit surface area ($\dot{M}_{\bigstar}/R^{\rm 2}_{\rm \bigstar}$, bottom) as a function of the maximum absolute radial magnetic field on the stellar surface ($|B_{\rm R}|^{\rm max}$). We see a trend where $\overline {AS}$ and $\dot{M}_{\bigstar}/R^{\rm 2}_{\rm \bigstar}$ increase with increasing magnetic field strength.
We fit this trend to a power law by applying the bootstrap method used to derive this parameter as a function of the average radial magnetic field (similar to the procedure used in Sect.~\ref{sec:3.1}, Eqs.~\ref{eq3} and \ref{equ4}).

\begin{equation}
    \log \overline{AS_{\rm R}} = (0.44 \pm 0.05) \log |B_{\rm R}|^{\rm max} + (0.54 \pm 0.08)\label{eq9}
\end{equation}

\begin{equation}
    \log \dot{M}_{\rm \bigstar} / R^{\rm 2}_{\bigstar} = (0.83 \pm 0.07) \log |B_{\rm R}|^{\rm max} - (0.48 \pm 0.10)\label{equ10}
\end{equation}

% Don't change these lines
\bsp	% typesetting comment
\label{lastpage}
\end{document}